\documentclass[epj]{svjour}
\usepackage{graphics}
\begin{document}

%
%
	
\title{Effects of next-to-leading order DGLAP evolution on generalized parton distributions of the proton and deeply virtual Compton scattering at high energy}

\author{Hamzeh Khanpour\inst{1,2} 
\and
Muhammad Goharipour\inst{2}
\and
Vadim Guzey\inst{3}	
}	
\offprints{}	
	
\institute{
Department of Physics, University of Science and Technology of Mazandaran, P.O.Box 48518-78195, Behshahr, Iran \and
School of Particles and Accelerators, Institute for Research in Fundamental Sciences (IPM), P.O.Box 19395-5531, Tehran, Iran \and
National Research Center ``Kurchatov Institute'', Petersburg Nuclear Physics Institute (PNPI), Gatchina, 188300, Russia}
\date{Received: date / Revised version: date}
%
%
\abstract{
We studied the effects of NLO $Q^2$ evolution of generalized parton distributions (GPDs) using 
the aligned-jet model for the singlet quark and gluon GPDs at an initial evolution scale.
We found that the skewness ratio for quarks is a slow logarithmic function of $Q^2$ reaching 
$r^S=1.5-2$ at $Q^2=100$ GeV$^2$ and $r^g \approx 1$ for gluons in a wide range of $Q^2$. 
Using the resulting GPDs, we calculated the DVCS cross section on the proton in NLO pQCD
and found that this model in conjunction with modern parameterizations of proton PDFs (CJ15 and CT14)
provides a good description of the available H1 and ZEUS data in a wide kinematic range.
}



\maketitle


\section{Introduction}\label{sec:Intro}

Generalized parton distributions (GPDs) have become a familiar and standard tool of Quantum Chromodynamics (QCD) describing the
response of hadronic targets in various hard exclusive processes~\cite{Mueller:1998fv,Radyushkin:1997ki,Ji:1996nm,Ji:1998pc,Goeke:2001tz,Belitsky:2001ns,Diehl:2003ny,Belitsky:2005qn}. 
GPDs can be rigorously defined in the framework of QCD collinear factorization for hard exclusive processes~\cite{Collins:1998be,Collins:1996fb}, which allows one to access universal, i.e., process-independent, GPDs in such 
processes as deeply virtual Compton scattering (DVCS) $\gamma^{\ast} + T \to \gamma + T$, timelike Compton scattering (TCS)
 $\gamma + T \to \gamma^{\ast} + T$, exclusive meson production by longitudinally polarized photons 
 $\gamma_L^{\ast} + T \to M + T$, and, recently, photoproduction of heavy ($J/\psi$, $\Upsilon$) vector mesons  
 $\gamma +T \to V+T$~\cite{Ivanov:2004vd,Jones:2015nna}.
GPDs contain information on the hadron structure in QCD, which is hybrid of that encoded in usual parton distributions and elastic form factors. In particular, GPDs describe the distributions of quarks and gluons in hadrons in terms of two light-cone momentum fractions and the position in the transverse plane. Also, GPDs are involved in the hadron spin decomposition in terms of the helicity and orbital motion contributions of quarks and gluons~\cite{Ji:1998pc,Goeke:2001tz,Belitsky:2001ns,Diehl:2003ny,Belitsky:2005qn}, and
carry information on the spatial distribution of forces experienced
by partons inside hadrons~\cite{Polyakov:2002yz}.

GPDs are essentially non-perturbative quantities, which cannot be calculated from the first principles apart from first
Mellin moments in special cases in lattice QCD~\cite{Hagler:2007xi,Alexandrou:2013joa}.
At the same time, evolution of GPDs with an increase of the resolution scale $Q^2$ is predicted by the QCD
Dokshitzer--Gribov--Lipatov--Altarelli--Parisi (DGLAP) evolution equations modified to the case of GPDs, 
which are presently known to the next-to-leading order (NLO) accuracy~\cite{Freund:2001hm,Freund:2001rk,Freund:2001hd}. 
Therefore, one of directions of phenomenological studies of GPDs is to determine the non-perturbative input for these evolution equations. After early studies of GPDs using various dynamical models of the nucleon structure~\cite{Ji:1997gm,Petrov:1998kf,Tiburzi:2001ta,Scopetta:2003et,Tiburzi:2004mh,Mineo:2005qr,Pobylitsa:2002vw,Freund:2002qf}, one currently focuses on parameterizations of GPDs, 
which are determined from fitting the available data. The two main contemporary approaches include the flexible parameterization based on the conformal expansion of GPDs~\cite{Kumericki:2007sa,Kumericki:2009uq,Lautenschlager:2013uya,Kumericki:2016ehc} and global fits of GPDs~\cite{Guidal:2008ie,Guidal:2010ig,Guidal:2013rya,Berthou:2015oaw}, which use the double distribution (DD) model~\cite{Radyushkin:1998es,Radyushkin:1998bz,Musatov:1999xp,Guidal:2004nd,Radyushkin:2013hca} in the Vanderhaeghen--Guichon--Guidal (VGG) framework, see details in~\cite{Guidal:2013rya}.
One should also mention a pioneering study of global QCD fits of GPDs within the neural network approach~\cite{Kumericki:2011rz}.

The mentioned above analyses present only a partial, model-dependent picture of GPDs in a limited kinematic range. For further progress, it is important to perform a systematic QCD analysis of evolution of GPDs and cross sections of hard exclusive processes involving them.
It will enable one to separate the effects of non-perturbative input GPDs from the perturbative DGLAP evolution and 
help to explore the possibility to use the data on hard exclusive reactions at high energies for constraining GPDs, 
see, e.g.~\cite{Aschenauer:2013hhw}.

In this paper, we calculate the effect of next-to-leading (NLO) QCD evolution on quark and gluon GPDs of the proton using the 
brute-force evolution method of~\cite{Freund:2001hm,Freund:2001rk,Freund:2001hd} and the physical model 
for input GPDs, which is motivated by the the aligned-jet model~\cite{Freund:2002qf}. 
Using the obtained results, we calculate the DVCS cross section on the proton in NLO QCD and compare it to the available HERA data.
We find that our approach provides a good description of the DVCS data over a wide kinematic range, including most 
of the data from H1 and ZEUS collaborations for the unpolarized proton target.

\section{Aligned-jet model for GPDs and QCD evolution effects}\label{sec:Modeling-of-GPDs}

\subsection{Input GPDs}\label{sec:Input-GPDs}

The aligned-jet model (AJM)~\cite{Bjorken:1973gc,Frankfurt:1988nt} for photon--hadron interactions at high energies is based on the general observation that 
in the target rest frame, the incoming photon first fluctuates into quark-antiquark configurations, which then interact with the target. For the photon virtualities $Q^2={\cal O}$(few)~GeV$^2$, 
the $q \bar{q}$ pair (dipole) is characterized by a small relative transverse momentum (hence the name aligned-jet), 
the invariant mass of the order of $Q^2$, the asymmetric sharing of the photon's light-cone momentum, and 
the dipole--nucleon cross section, which has the magnitude typical for hadron--nucleon cross sections.
Note that in QCD, this parton picture is complimented by the gluon emission and the contribution of quark-antiquark dipoles with large
relative transverse momenta, which become progressively important as $Q^2$ is increased; see the discussion in Ref.~\cite{Frankfurt:2013ria}.

In the AJM model, one obtains for the ratio of the imaginary parts of DVCS and DIS amplitudes at $Q^2=1-3$ GeV$^2$,
$R=\Im {\cal T}_{\rm DVCS}/\Im {\cal T}_{\rm DIS}=2.5-3.5$~\cite{Freund:2002qf,Frankfurt:1997at}, which agrees nicely with the values of $R$ extracted from the HERA data~\cite{Schoeffel:2007dt}.
This in turn means that the effect of skewness of the singlet quark GPDs in the DGLAP region of $X \geq \zeta$ can be neglected
($X$ is the light-cone momentum fraction of the target in the initial state carried by the interacting parton;
$\zeta$ is the momentum fraction difference between the two interacting partons).
This observation is also supported by the analysis of Ref.~\cite{Belitsky:2001ns}, which showed that the good description of
the high-energy HERA data on the DVCS cross section on the proton can be achieved with the forward parton distribution model
for the singlet quark GPDs~\cite{Guichon:1998xv,Vanderhaeghen:1999xj}, i.e., with the $\delta$-function-like profile in the 
DD model for sea quark GPDs.

In general, modeling and parametrization of GPDs is a non-trivial task since GPDs should satisfy several general constraints:
GPDs reduce to usual parton distributions functions (PDFs) in the forward limit; 
integration of GPDs over the momentum fraction gives the 
corresponding elastic form factors; as a consequence of Lorentz invariance, 
Mellin moment of GPDs are finite-order polynomials in even powers of the skewness $\eta=\zeta/(2-\zeta)$ 
(the property of polynomiality); GPDs obey positivity bounds expressed an inequalities involving GPDs and usual PDFs.
While the first three properties can be naturally implemented in momentum representation of GPDs, positivity is most naturally
derived in coordinate representation. Hence, it is an outstanding challenge to propose a practical model of GPDs
satisfying all these constraints. (Naturally, field-theoretical approaches based on perturbative diagrams will automatically
lead to GPDs satisfying all the constrains~\cite{Pobylitsa:2002vw}, but they have little usefulness for GPD phenomenology.)

Starting from a model for GPDs in the DGLAP region of $X \geq \zeta$, there is no unique and simple way to reconstruct GPDs in the entire range of $X$. For instance, the method proposed in~\cite{Freund:2002qf,Schoeffel:2007dt} does not guarantee polynomiality for higher moments of 
GPDs and conflicts with dispersion relations (DR) for the real and imaginary parts of the DVCS amplitude~\cite{Diehl:2007jb}.
In principle, GPDs with the correct forward limit and satisfying the property of polynomiality can be constructed using the so-called
Shuvaev transform~\cite{GolecBiernat:1999ib,Guzey:2005ec,Guzey:2006xi}. However, this method is usually associated with the leading order (LO) phenomenology and 
also brings certain skewness dependence of GPDs in the DGLAP region.
Similarly, the flexible parameterization of GPDs based on the conformal expansion~\cite{Kumericki:2007sa,Kumericki:2009uq,Lautenschlager:2013uya,Kumericki:2016ehc} contains the skewness effect of GPDs in the DGLAP region and also corresponds to model-dependent parton distributions in the forward limit.

In this work, to simultaneously have the forward-like GPDs in the DGLAP region and circumvent the aforementioned problem with 
polynomiality, we take forward-like GPDs for all $X$ and add the so-called $D$-term~\cite{Polyakov:1999gs}, which
has support only in the Efremov--Raduyshkin--Brodsky--Lepage (ERBL) region of $|X| \leq \zeta$. Specifically, 
we use the following model for the singlet quark (one sums over quark flavors $q$) and gluon GPDs
at $t=0$ at the initial scale of $\mu_0$:
\begin{eqnarray}
&&(1-\zeta/2) \, H^S(X,\zeta,t=0,\mu_0) = \nonumber\\
&&\left\{\begin{array}{ll}
\sum_{q} \left[q(x,\mu_0)+\bar{q}(x,\mu_0)\right]+D^S\left(x/\eta\right)\theta(\zeta-X) \,, \quad & X > \zeta/2 \\
-\sum_q \left[q(x,\mu_0)+\bar{q}(x,\mu_0)\right]-D^S\left(x/\eta\right)\theta(\zeta-X) \,, \quad & X < \zeta/2 
\end{array} \right. \nonumber \\
&&(1-\zeta/2) \, H^g(X,\zeta,t=0,\mu_0) = |x|g(|x|,\mu_0) \,, 
\label{eq:input}
\end{eqnarray}
where $x=(X-\zeta/2)/(1-\zeta/2)$ and $\eta=\zeta/(2-\zeta)$;
$q(x,\mu)$ and $g(x,\mu)$ are the quark and gluon parton distribution functions (PDFs), respectively.
Note that since we explicitly introduced antiquark GPDs, it is sufficient to consider only non-negative $X \geq 0$. 
Also, we assume that similar relations hold for separate quark flavors $q$, i.e.,
\begin{eqnarray}
&&(1-\zeta/2) \, H^{q+\bar{q}}(X,\zeta,t=0,\mu_0) = \nonumber \\
&&\left\{\begin{array}{ll}
q(x,\mu_0)+\bar{q}(x,\mu_0)+\frac{1}{n_f} D^S\left(x/\eta\right)\theta(\zeta-X) \,, \quad & X > \zeta/2 \\
-\left[q(x,\mu_0)+\bar{q}(x,\mu_0)\right]-\frac{1}{n_f}D^S\left(x/\eta\right)\theta(\zeta-X) \,, \quad & X < \zeta/2 
\end{array} \right.  \,,  \nonumber \\
\label{eq:input_2}
\end{eqnarray}
where $n_f$ is the number of active quark flavors. 
Thus, our model does not assume the flavor symmetry
of quark GPDs.

We should stress here that GPDs have to be continuous at $X = \zeta$.
In addition, GPDs have to satisfy the correct symmetries around the midpoint of the ERBL region, $X=\zeta/2$.
As follows from general properties of GPDs, the singlet quark GPDs $H^S(X,\zeta,t=0,\mu_0)$ is antisymmetric in the ERBL region around the $X=\zeta/2$ point, while the gluon GPD 
$H^g(X,\zeta,t=0,\mu_0)$ is symmetric in the ERBL region;
these constraints are implemented in Eqs.~(\ref{eq:input}) and (\ref{eq:input_2}).

The function $D^S\left(x/\eta\right)$ is the singlet quark $D$-term~\cite{Polyakov:1999gs}, which can be expanded in terms of  
odd Gegenbauer polynomials $C_n^{3/2}$ in the following form~\cite{Kivel:2000fg}:
\begin{eqnarray}
&& D^S(z,\mu_0)= \nonumber\\
&& 2(1-z^2)[d_1 C_1^{3/2}(z)+d_3 C_3^{3/2}(z)+d_5 C_5^{3/2}(z)] \,.
\label{eq:D-term}
\end{eqnarray}
The coefficients $d_1$, $d_3$ and $d_5$ were estimated in the chiral quark soliton model at $\mu_0=0.6$ GeV in Ref.~\cite{Petrov:1998kf}:
$d_1=-4$, $d_3=-1.2$, and $d_5=-0.4$. Note that due to the lack of numerical estimates, we neglected the possible gluon $D$-term
in Eq.~(\ref{eq:input}). In this case, $D^S(z,\mu)$ evolves in $\mu^2$ autonomously (without mixing) and its value for 
$\mu > \mu_0$ can be readily calculated.

In summary, our GPD model in Eqs.~(\ref{eq:input}) and (\ref{eq:input_2}) corresponds to the correct forward limit, 
satisfies polynomiality (the $D$-term satisfies polynomiality by construction), and obeys positivity bounds in 
the DGLAP region in
the small-$\xi$ and $t=0$ limit (all positivity bounds discussed in the literature are for the DGLAP region, see
Ref.~\cite{Diehl:2003ny}). 
Indeed, neglecting the kinematically-suppressed contribution of the GPD $E$, 
the positivity bound for the quark GPDs reads~\cite{Diehl:2003ny}:
\begin{equation}
(1-\xi^2)[H^q(x,\eta,t=0)]^2 \leq q(x_{\rm in}) q(x_{\rm out}) \,,
\label{eq:positivity}
\end{equation}
where $H^q(x,\eta)=(1-\zeta/2) H^q(X,\zeta)$; $x_{\rm in}=(x+\xi)/(1+\xi)=X$ and $x_{\rm out}=(x-\xi)/(1-\xi)=(X-\zeta/2)/(1-\zeta/2)$.
Assuming that $q(x) \propto 1/x^{\alpha}$ for small $x$, where $0 \leq \alpha \leq 1$, Eq.~(\ref{eq:positivity}) is trivially
satisfied with our GPD model of Eq.~(\ref{eq:input_2}).

By construction, see Eq.~(\ref{eq:input}), in the middle of the ERBL region at $x=X-\zeta/2=0$, our singlet quark GPDs become
singular and the gluon GPD vanishes. Being a natural artifact of our model imposing the correct GPD symmetry in the ERBL problem,
it does not violate general principles of GPDs, does not conflict with factorization for amplitudes of hard exclusive processes, 
and does not lead to singularities of the DVCS amplitude. Since the main goal of our work is to study the effects of NLO $Q^2$ evolution of GPDs in conjunction with different baseline
PDFs, the simple model of Eq.~(\ref{eq:input}) should suffice.

Note that in this work, we focus on the quark singlet $\sum_q(q + \bar{q})$ and gluon GPDs: 
valence (non-singlet) quark GPDs do not mix with singlet quark and gluon GPDs under the DGLAP evolution and do not 
appreciably contribute to the DVCS amplitude at high energies.

\subsection{NLO $Q^2$ evolution of GPDs and error analysis}\label{sec:evolution-of-GPDs}

The determination of parton distribution functions (PDFs) has always been one of the important ingredients for theory predictions. In this respect, more accurate PDFs play an important role in understanding of hadronic properties and the structure of the 
nucleon~\cite{Ball:2017nwa,Hou:2016nqm,Harland-Lang:2014zoa,Harland-Lang:2016yfn}. From past to present, our knowledge of PDFs has been developed both theoretically and computationally. 
However, results of various groups lead to different predictions of physical observables. As we know, 
GPDs are quantities that are related to the PDFs in the forward limit and in many phenomenological approaches. 
To investigate the impact of different PDFs on the GPDs and their evolution, 
we calculate the effect of next-to-leading order (NLO) DGLAP evolution equations
modified to the case of GPDs using the formalism of~\cite{Freund:2001hm,Freund:2001rk,Freund:2001hd}
and the input GPDs of Eqs.~(\ref{eq:input}).
(The early results on leading order (LO) $Q^2$ evolution of GPDs were presented in Refs.~\cite{GolecBiernat:1999ib,Frankfurt:1997ha}.) 
Perturbation theory predicts the evolution of GPDs and, hence, they depend on the factorization scale, $\mu^2$.
Anomalous dimensions and the kernels at NLO accuracy in pQCD can be found in Refs.~\cite{Belitsky:1999hf,Belitsky:1999fu,Belitsky:1998vj,Belitsky:1998gc,Belitsky:1999gu}.

For the forward PDFs, we used CT14~\cite{Dulat:2015mca} and the new CTEQ-Jefferson Lab (CJ15) analysis~\cite{Accardi:2016qay}. 
To study the impact of PDF uncertainties on the GPD evolutions and DVCS cross sections, we include the uncertainties of CT14 and CJ15 PDFs in the calculations of the evolution and also in the DVCS cross sections. 
In this respect, we note that both CT14 and CJ15 are PDF sets with Hessian PDF eigenvector error sets. In this situation, the theoretical uncertainties of PDFs themselves
and also any physical quantity related to them, such as the GPDs and DVCS cross sections considered here, can be obtained as usual using the 56 and 48 error sets of the CT14 and CJ15 parametrizations, respectively. To this aim, we must first calculate our desired quantity with various error sets. Then, we can compute the deviations from the central result and
so the contribution to the size of the upper and lower
errors through the following relations~\cite{Pumplin:2001ct,Nadolsky:2008zw}:
\begin{eqnarray}
\delta^{+} X &=& \sqrt{\sum_{i} [\max (X^{(+)}_{i} - X_{0}, X^{(-)}_{i} - X_{0}, 0)]^{2}}\,   \nonumber  \\
\delta^{-} X &=& \sqrt{\sum_{i} [\max (X_{0} - X^{(+)}_{i}, X_{0} - X^{(-)}_{i}, 0)]^{2}}\,.
\label{error}
\end{eqnarray}
The last point that should be noted here is the confidence region considered for estimating
error bands 
since different PDF analyses typically utilize different criteria for estimating PDF errors. The CJ15 PDF sets have been provided with 90 \% C.L uncertainties considering standard 
tolerance criterion $\Delta\chi^2$ = 2.71, while CT14 use a tolerance criterion as $\Delta\chi^2$ = 100 with the same confidence level~\cite{Accardi:2016ndt}. 
In this work, we display the CT14 and CJ15 errors on GPDs and 
DVCS cross sections for 90\% C.L. region, so that the tolerance used for CJ15 PDFs
be matched with CT14, in order to have a reasonable comparison.

\begin{figure*}[htb]
	\begin{center}
		\vspace{0.50cm}
		\resizebox{0.52\textwidth}{!}{\includegraphics{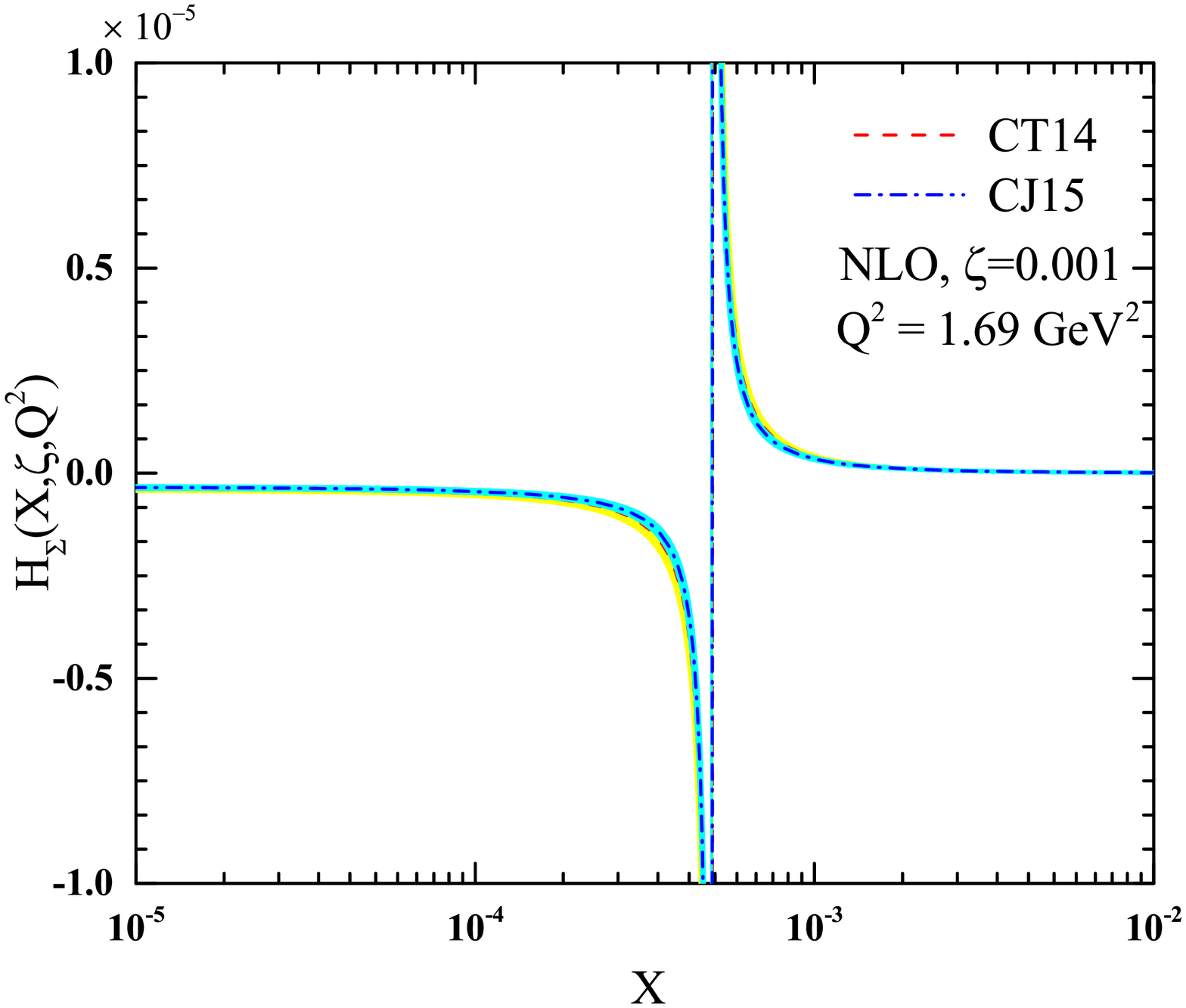}}   
		\hspace{-10mm}
		\resizebox{0.52\textwidth}{!}{\includegraphics{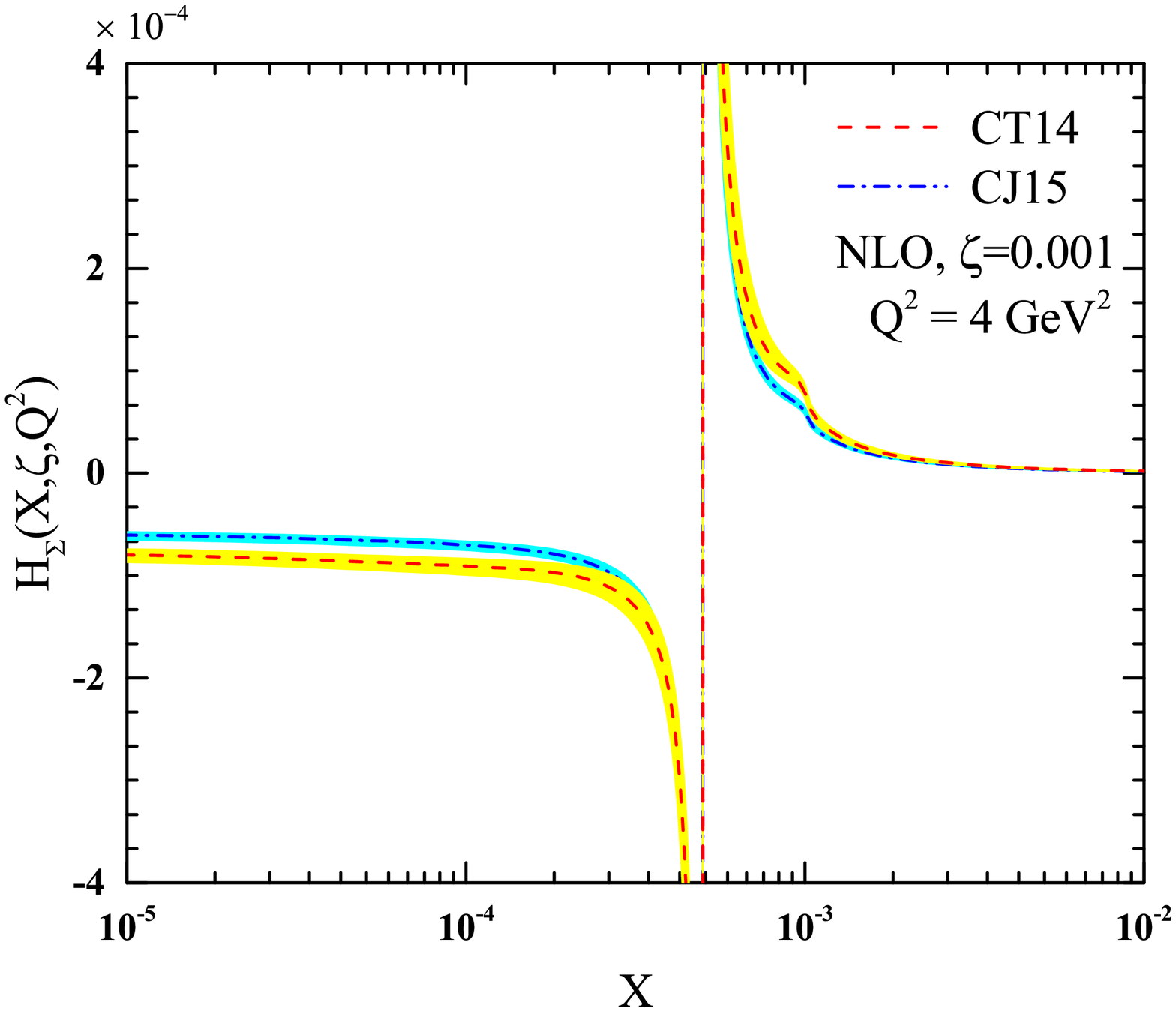}}   
		\resizebox{0.52\textwidth}{!}{\includegraphics{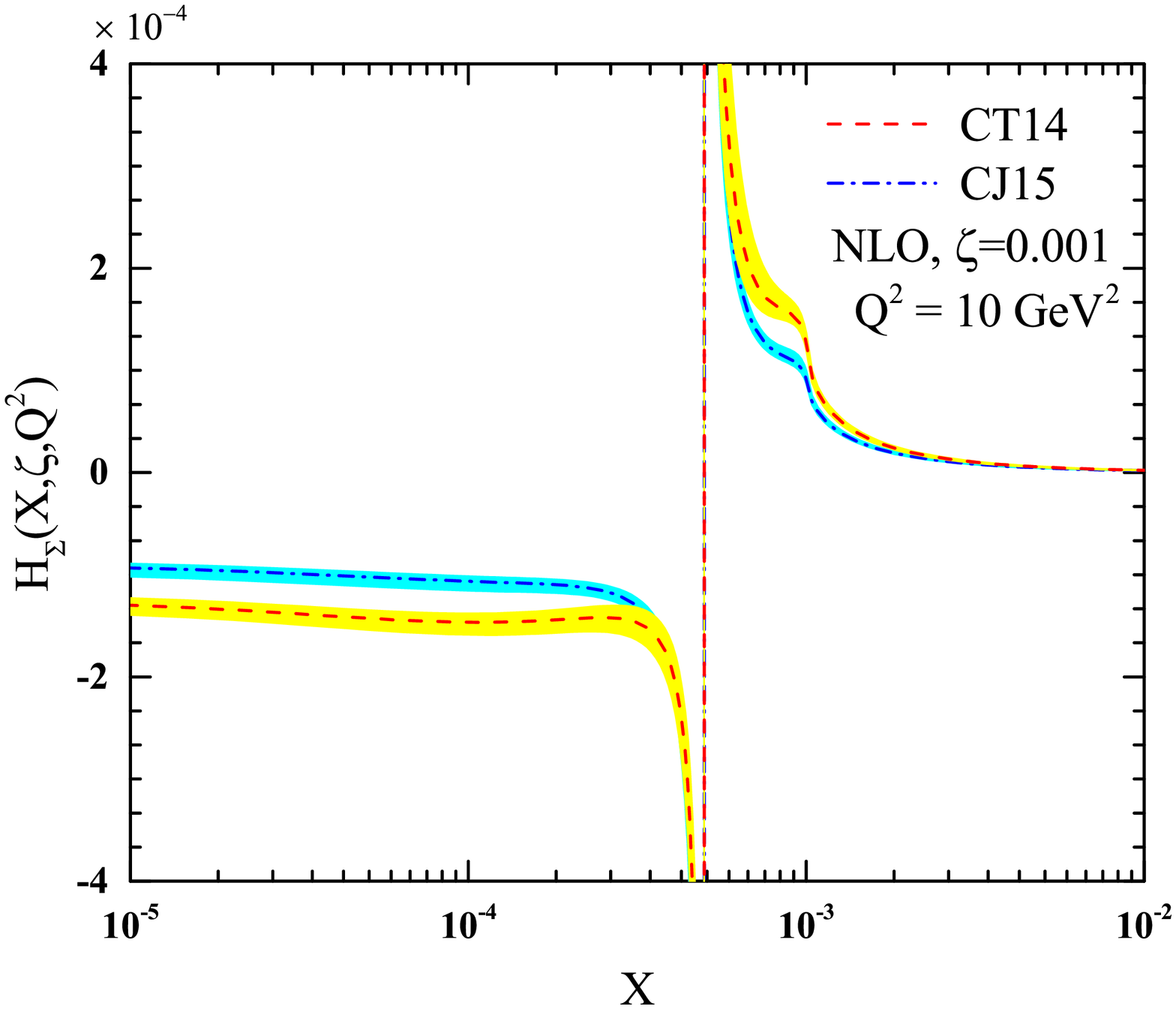}}   
		\hspace{-10mm}
		\resizebox{0.52\textwidth}{!}{\includegraphics{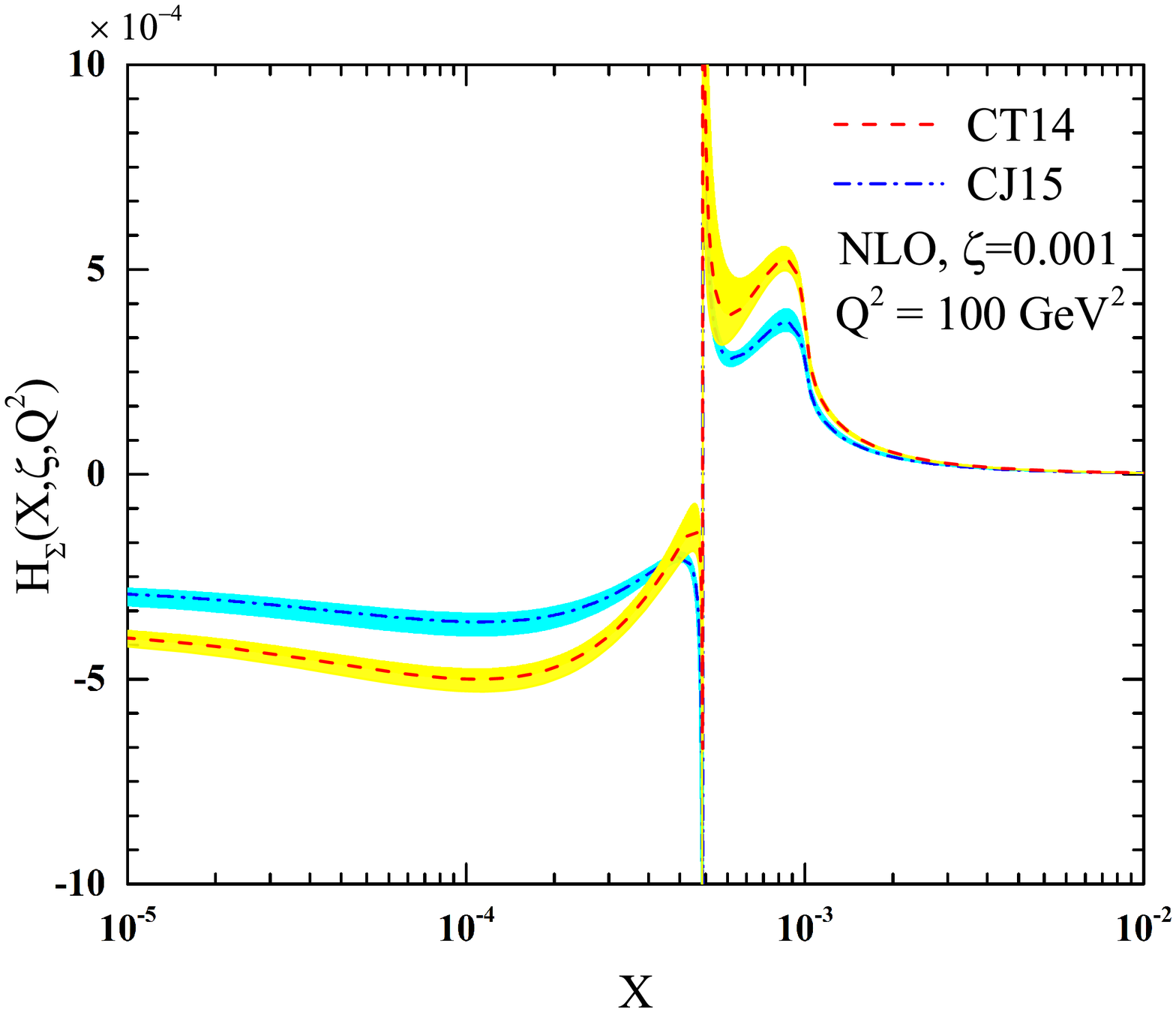}}   
		\caption{(Color online) The singlet quark GPD $H^S(X, \zeta, t=0,Q^2)$ as a function of $X$ at $\zeta = 0.001$ and 
		$Q^2 = 1.69$, 4, 10 and 100 GeV$^2$. The GPDs are calculated using the input of Eq.~(\ref{eq:input}) 
		with the CT14~\cite{Dulat:2015mca} and CJ15~\cite{Accardi:2016qay} parameterizations of PDFs and NLO $Q^2$ evolution
		 for GPDs.}
		\label{fig:H_s_GPDs}
		\end{center}
\end{figure*}

\begin{figure*}[htb]
	\begin{center}
		\vspace{0.50cm}
		\resizebox{0.52\textwidth}{!}{\includegraphics{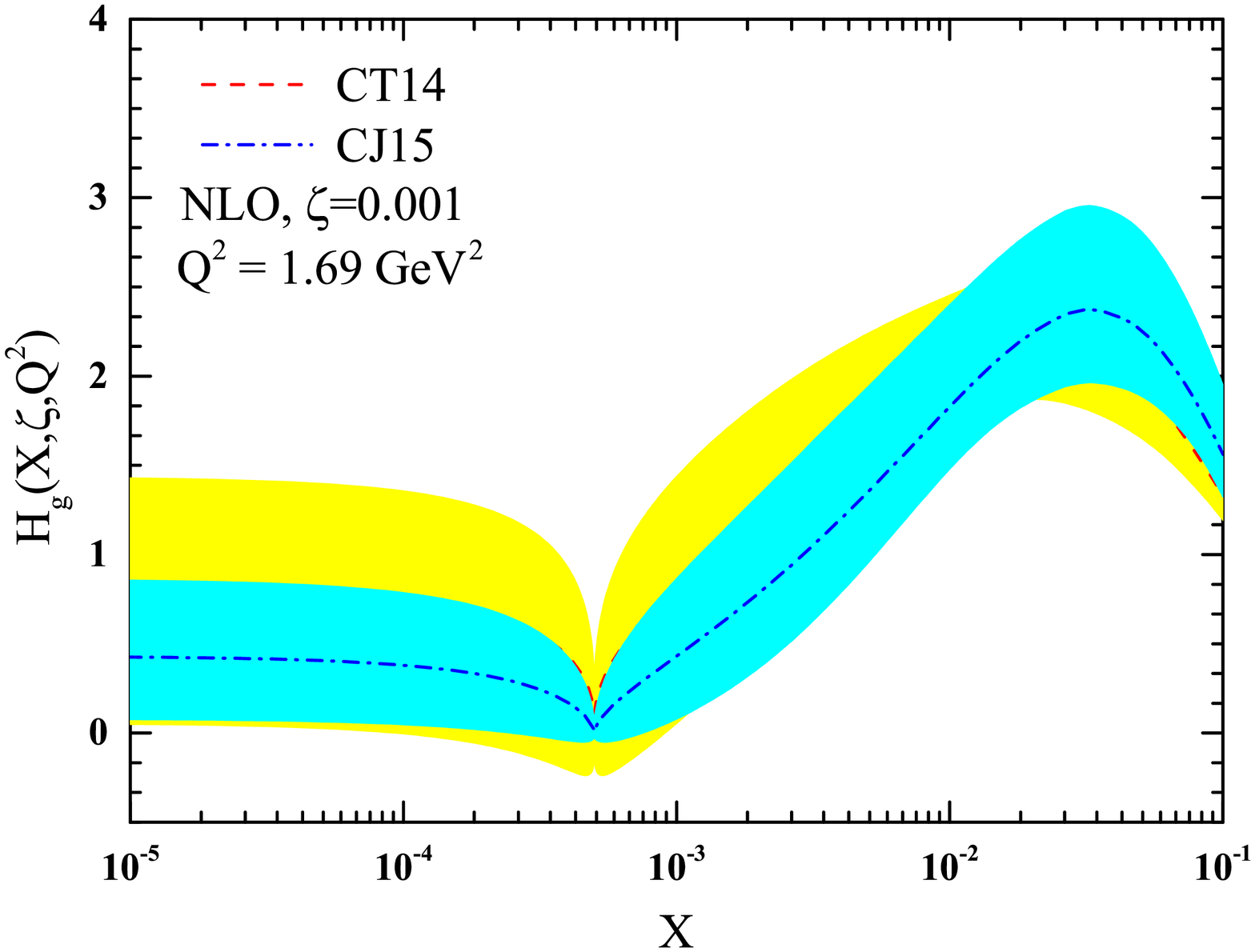}}   
		\hspace{-10mm}
		\resizebox{0.52\textwidth}{!}{\includegraphics{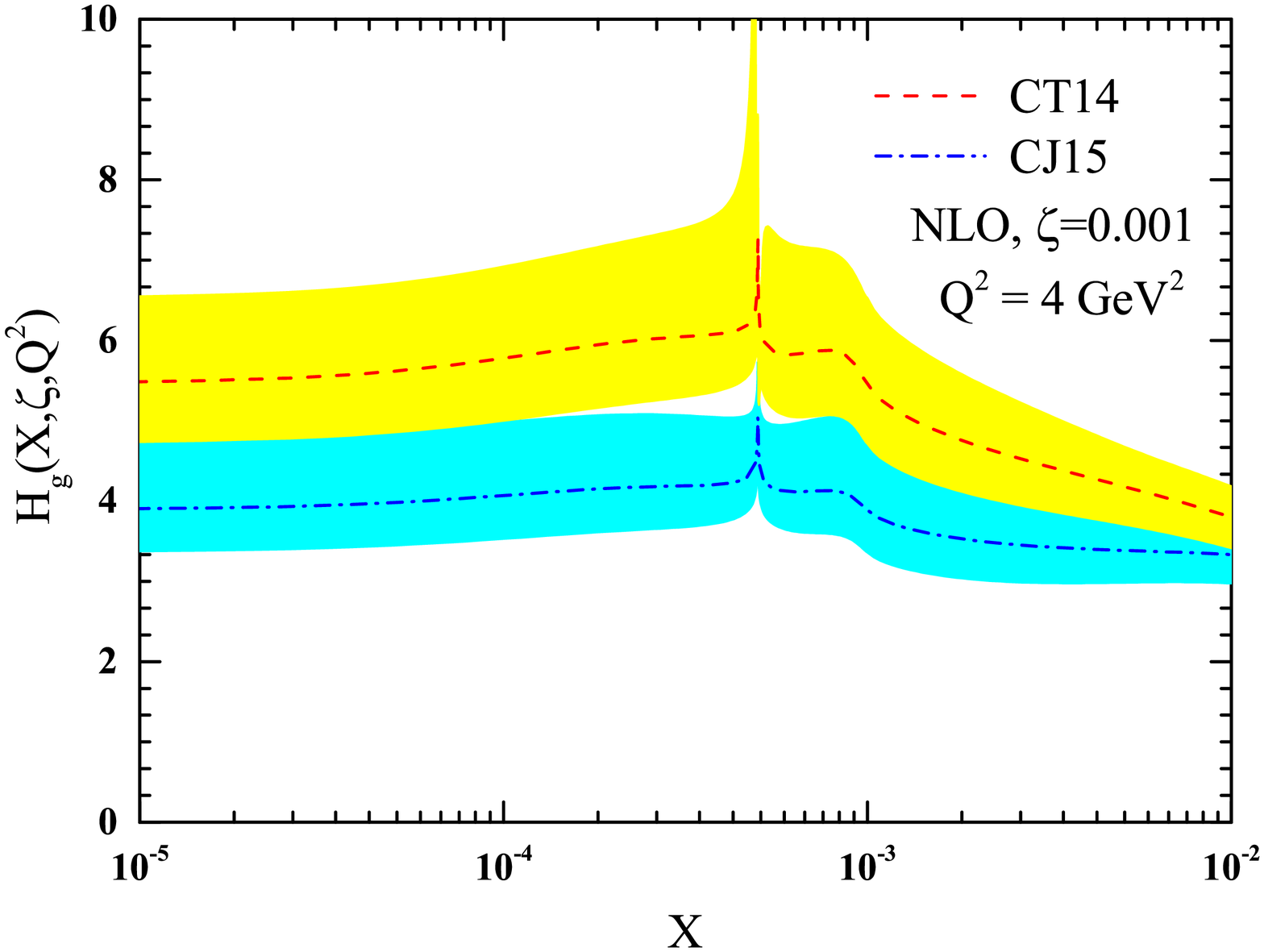}}   
		\resizebox{0.52\textwidth}{!}{\includegraphics{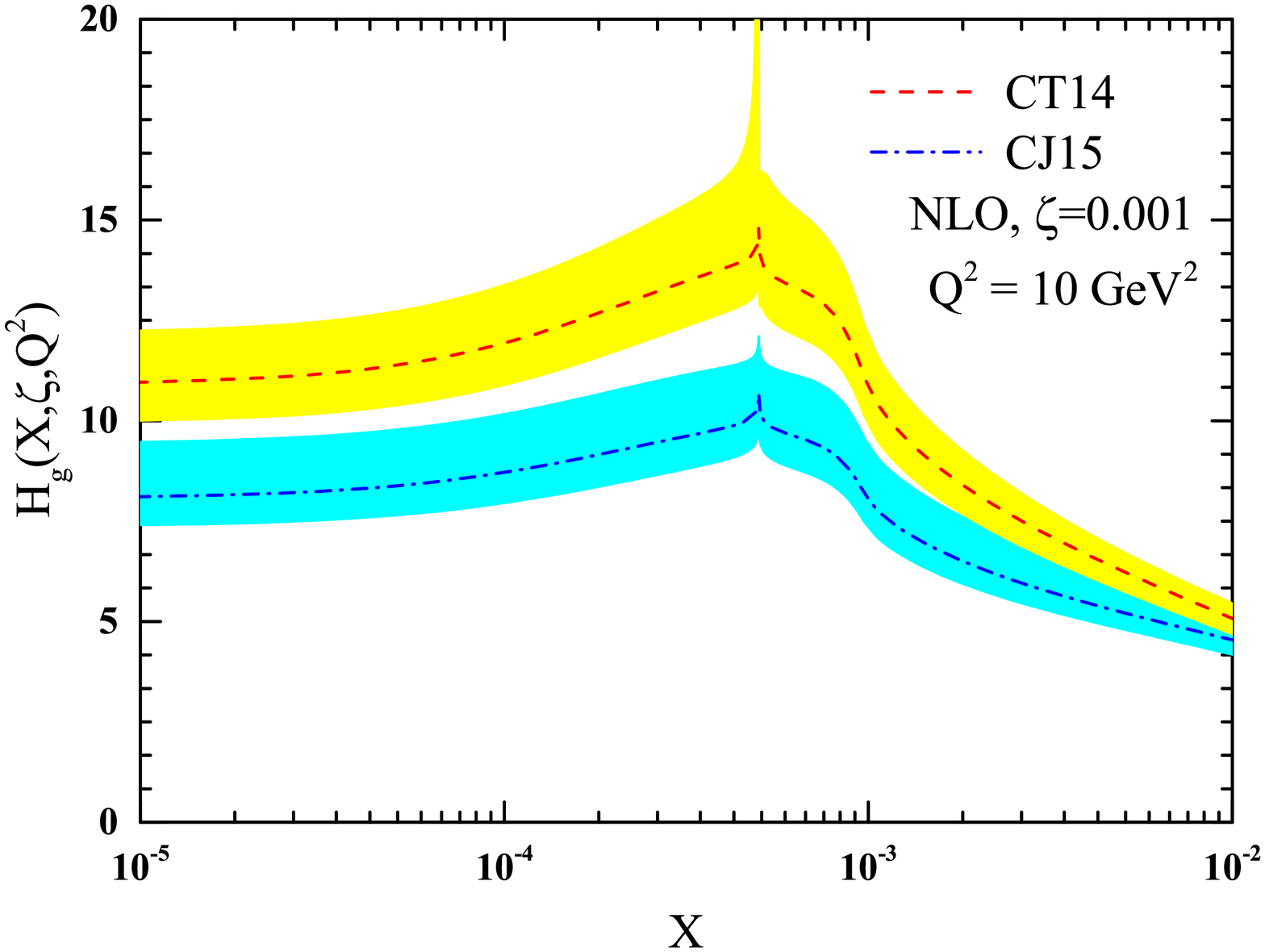}}   
		\hspace{-10mm}
		\resizebox{0.52\textwidth}{!}{\includegraphics{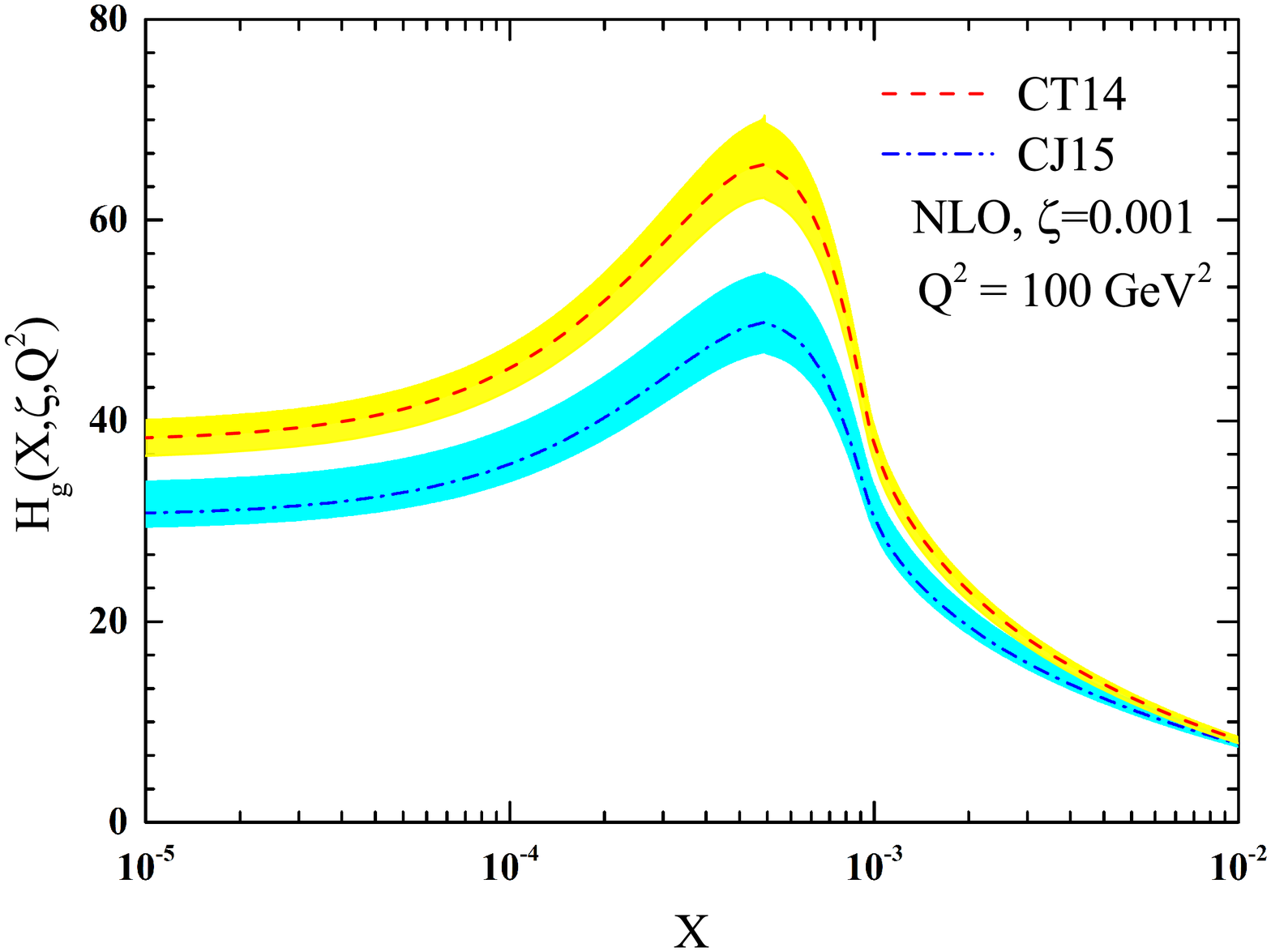}}   
		\caption{(Color online) The gluon GPD $H^g(X, \zeta, t=0,Q^2)$ as a function of $X$ at $\zeta = 0.001$ and 
		$Q^2 = 1.69$, 4, 10 and 100 GeV$^2$. See the caption of Fig.~\ref{fig:H_s_GPDs} for details.}
		\label{fig:H_g_GPDs}
	\end{center}
\end{figure*}

Figures~\ref{fig:H_s_GPDs} and \ref{fig:H_g_GPDs} show the results for the singlet quark GPD $H^S(X, \zeta, t=0,Q^2)$ and 
the gluon GPD $H^g(X, \zeta,t=0, Q^2)$, respectively, as a function of $X$ at $\zeta = 0.001$ and $Q^2 =1.69$, 4, 10, and 100 GeV$^2$. 
Note that $Q^2 =1.69$ GeV$^2$ is the input scale for CT14 and CJ15.
As can be seen from these figures, the $Q^2$ evolution pushes GPDs into the ERBL region of $X < \zeta$ as it should be.
The discontinuity of the quark singlet GPD at $X=\zeta/2$ is an artifact of our model (see the discussion 
in Sec.~\ref{sec:Input-GPDs}), which does not affect the physical observables.

In the quark singlet case, the difference between the predictions based on CT14 and CJ15 PDF is small, especially at lower values of the $Q^2$ resolution scale. At the same time, in the gluon channel the differences between the CT14 and CJ15 
predictions are sizable and exceed the associated uncertainties for large values of $Q^2$.
One should also note that the uncertainties of the resulting GPDs based on CT14 are larger than those for CJ15, which
is related to the large uncertainties of CT14 singlet distributions at small $x$.

Generally speaking, our results indicate that the GPD model of Eq.~(\ref{eq:input}) is sensitive to the input PDFs. 
Therefore, more accurate PDFs are very important for physical observables involving GPDs such as, e.g., DVCS cross sections. 
Conversely and optimistically, 
data on the DVCS cross section may provide new constraints for global QCD analysis of PDFs.
Our study makes it clear that using more recent version of PDFs and proper scale dependence in our GPDs model describes 
DVCS data over a large kinematical range.  

\subsection{Effect of skewness}\label{sec:Effect-of-skewness}

\begin{figure*}[htb]
	\begin{center}
		\resizebox{0.52\textwidth}{!}{\includegraphics{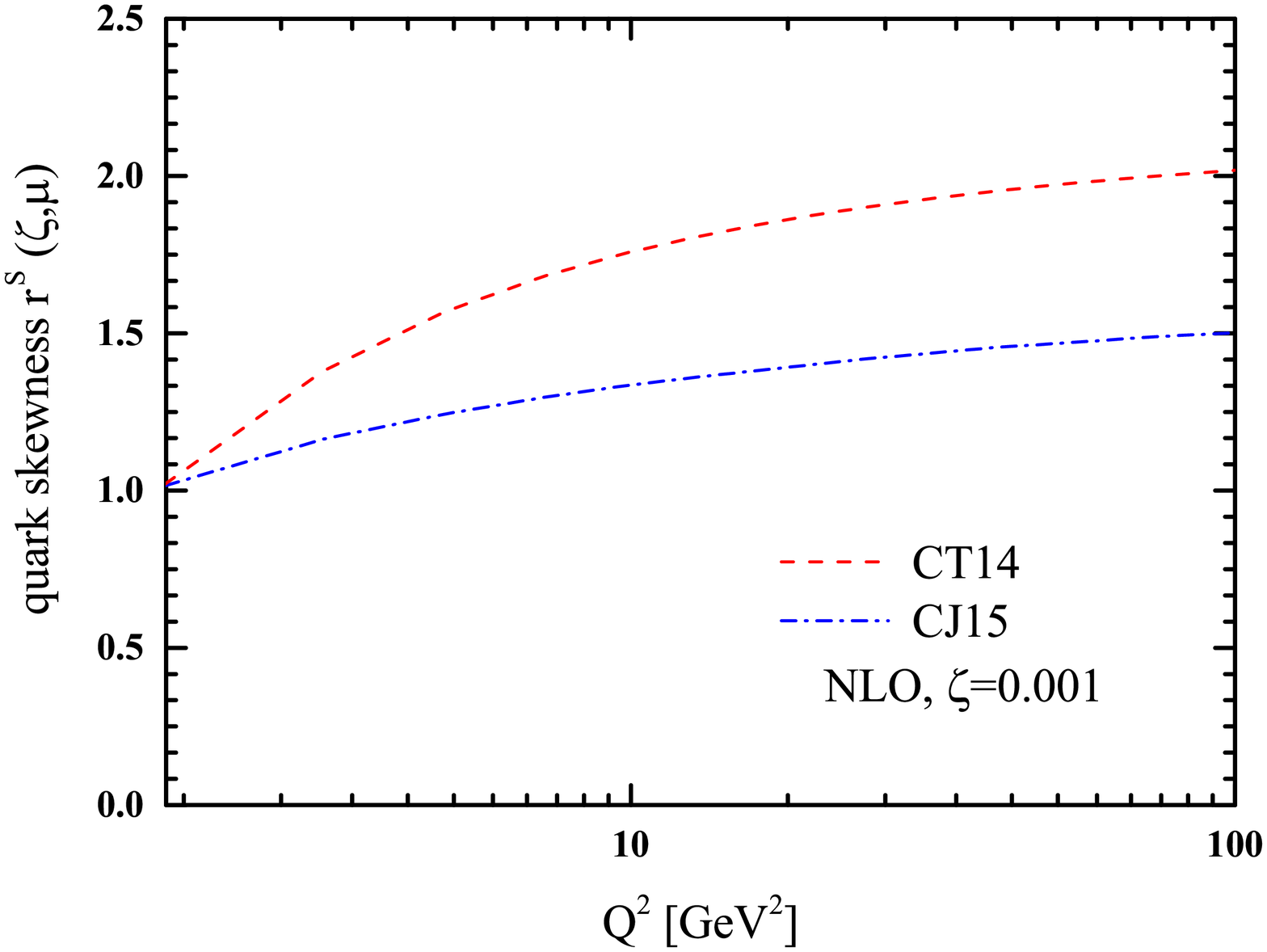}}\label{fig:r-Sigma}   
		\hspace{-10mm}
		\resizebox{0.52\textwidth}{!}{\includegraphics{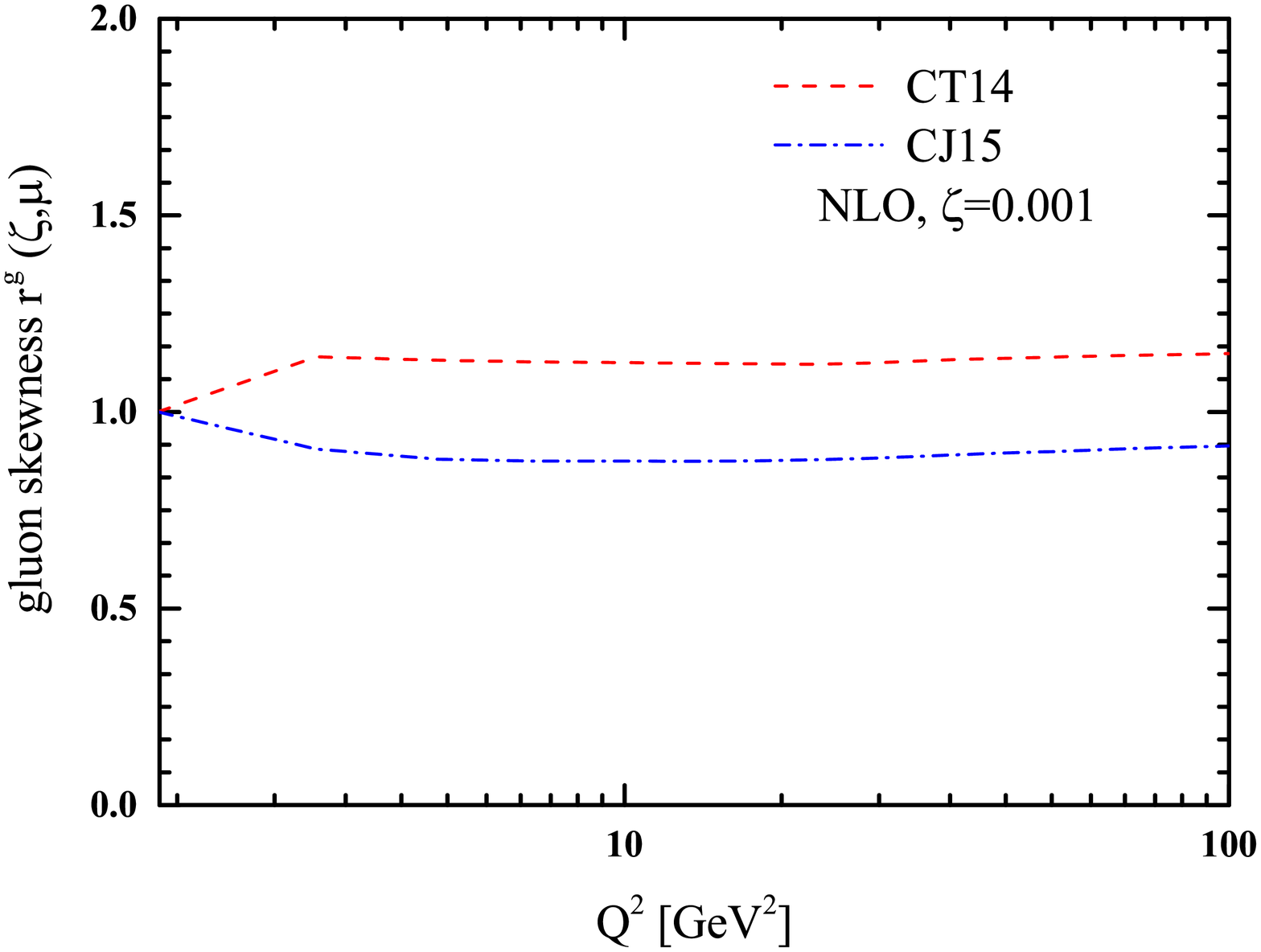}}\label{fig:r-G}   
		\caption{(Color online) The quark and gluon skewness ratios 
		$r^S(\zeta,\mu)$ (left) and $r^g(\zeta,\mu)$ (right) as functions $Q^2=\mu^2$ at $\zeta=0.001$. }
		\label{fig:r}
	\end{center}
\end{figure*}

For phenomenological applications of GPDs, it is important to discuss the so-called skewness factor, which 
describes the connection between GPDs and PDFs and parametrizes the deviation of GPDs from PDFs. 
To quantify this effect, it is convenient to introduce the following ratios of quark and gluon GPDs and PDFs~\cite{Kumericki:2009uq}:
\begin{eqnarray}
r^S(\zeta,\mu) &=& \frac{(1-\zeta/2)H^S(\zeta,\zeta,t=0,\mu)}{\sum_q \left[q(\zeta/(2-\zeta), \mu)+{\bar q}(\zeta/(2-\zeta), \mu)\right]} \,, \nonumber \\
r^g(\zeta,\mu) &=& \frac{(1-\zeta/2)H^g(\zeta,\zeta,t=0,\mu)}{\zeta/(2-\zeta)g(\zeta/(2-\zeta), \mu)} \,.
\label{eq:r-small}
\end{eqnarray}
Our results for $r^S(\zeta,\mu)$ and $r^g(\zeta,\mu)$ as functions $Q^2=\mu^2$ at $\zeta=0.001$ are 
shown in Fig.~\ref{fig:r}. One can see from the figure that both $r^S$ and $r^g$ are slow logarithmic functions of $Q^2$.
By construction, $r^S=r^g=1$ at the initial evolution scale of $Q^2=1.69$ GeV$^2$.
As $Q^2$ is increased, $r^S$ slowly increases up to $r^S \approx 1.5-2$ at $Q^2=100$ GeV$^2$, while $r^g$ stays at the 
level of unity for the studied range of $Q^2$.

These results agree with the predictions of the flexible GPD parameterization based on the conformal expansion, see
Fig.~7 of Ref.~\cite{Kumericki:2009uq}, except for $r^S$ at the input $Q^2=1.69$ GeV$^2$, where our result lies lower than that 
of~\cite{Kumericki:2009uq}.

\section{NLO pQCD predictions for the DVCS cross section and comparison to HERA data}\label{sec:dvcs}

\subsection{Evaluation of Compton form factors and DVCS amplitudes}\label{sec:CFFs}

The standard and well-tested way to access GPDs is the process of leptoproduction of a real photon, $e p \to e \gamma p$,
or deeply virtual Compton scattering (DVCS). At the photon level, the $\gamma^{\ast} p \to \gamma p$ DVCS differential cross section,
$d \sigma^{\rm DVCS}(W, t, Q^2)/dt$, is expressed in terms of the so-called Compton form factors (CFFs), which 
in the collinear factorization approach~\cite{Collins:1998be} are given as convolution of the perturbatively calculable hard scattering coefficient functions with the non-perturbative GPDs. In particular, at high energies the DVCS cross section is by far dominated by the GPD $H$ and its corresponding CFFs. For the flavor singlet contribution (for the quark singlet and gluon CFFs),
 one has in the symmetric notation:
\begin{eqnarray}\label{eq:CFFs}
{\cal F}^{S,g} (\xi, t, Q^2) = \int_{-1}^{+1} \frac{dx}{\xi} && \, C^{S,g}(x/\xi, Q^2/\mu^2, \alpha_s(\mu))  \\  \nonumber
&& \times H^{S,g} (x, \xi, t, \mu^2)  \,,
\end{eqnarray}
where $\xi = \zeta/(2 - \zeta)$;
$\mu$ is the factorization scale which is usually set equal to the photon virtuality $\mu^2 = Q^2$.
The explicit form for the coefficient functions $C$ can be found in Refs.~\cite{Mueller:2005nz,Kumericki:2007sa,Kumericki:2006xx} for the non-singlet and singlet cases.
For instance, for the quark singlet case, the QCD perturbation series reads:
\begin{eqnarray}\label{eq:C-QCD}
\frac{1}{\xi}  C^S(x/\xi, Q^2/\mu^2, \alpha_s(\mu)) = \frac{1}{\xi - x - i \epsilon} + {\cal O} (\alpha_s)  \,.
\end{eqnarray}
Hence, to the LO accuracy of pQCD and in leading-twist approximation, the DVCS scattering amplitude (CFF)
can be written as
\begin{equation}\label{eq:CFFs-LO}
{\cal F} (\xi, t, Q^2) = \sum_{q=u,d,s,...} e_q^2 \int_{-1}^{+1} \frac{dx}{\xi - x - i \epsilon}  H^{q+\bar{q}} (x, \xi, t, Q^2)  \,.
\end{equation}
The CFFs depend on $\xi$ (or equivalently on Bjorken $x_B$ or the invariant energy $W$), the momentum transfer $t$,
and $Q^2$ and, hence, can be extracted from DVCS experiments. 
Note that our model for the GPD initial conditions does not imply the flavor symmetry of quark GPDs.

Note that the purely electromagnetic Bethe-Heitler (BH) bremsstrahlung process leads to the same final state and interferes
with DVCS. However, at high energies (small values of $x_B$), the DVCS process dominates, which allows one to extract the DVCS cross section by subtracting of the BH contribution.
In addition to this, cuts on $Q^2$ and $W$ have been applied by H1 and ZEUS collaborations to enhance the contribution from DVCS process (see the following subsection).

Detailed analytic expressions for the DVCS and BH amplitudes squared and their interference are well-known and can be found in
Refs.~\cite{Kroll:2012sm,Freund:2003qs}. 
In our analysis we assume an exponential and factorized  $t$-dependence of the DVCS cross section, 
$e^{-b(Q^2)|t|}$, where	$b(Q^2) = a [ 1 - c \, \ln(Q^2/2 \, {\rm GeV}^{2}) ]$, with $a = 8 \, {\rm GeV}^{-2}$ and $c=0.15$~\cite{Freund:2002qf}. This simple parametrization agrees with the measurements of 
the $t$ dependence of the differential $\gamma^* p \to \gamma p$ cross section at HERA (see the following subsection).

\subsection{HERA DVCS data}\label{sec:DVCS-at-HERA}

Unpolarized DVCS on the proton has been measured in $e^\pm p$ collisions at HERA by the H1~\cite{Adloff:2001cn,Aktas:2005ty,Aaron:2007ab,Aaron:2009ac} and ZEUS~\cite{Chekanov:2003ya,Chekanov:2008vy} experiments.
The list of the DVCS experiments at HERA along with the measured observables, kinematic ranges and  corresponding references
is given in Table~\ref{table:DataSet}.

\begin{table*}
\begin{tabular}{ c | c c c c c }
\hline   \hline
Collaboration	&  Observables      &   $Q^2$ [GeV$^2$]  &     W [GeV] &     Year  &  Reference                \\     \hline    \hline
H1 & $\sigma_{\rm DVCS}(Q^2), \sigma_{\rm DVCS}(W)$   & 2-20    &  30-120  & 2001 &   \cite{Adloff:2001cn}     \\  
H1 & $\sigma_{\rm DVCS}(Q^2), \sigma_{\rm DVCS}(W)$   & 2-80    &  30-140  & 2005 &   \cite{Aktas:2005ty}      \\  
H1 & $\sigma_{\rm DVCS}(Q^2), \sigma_{\rm DVCS}(W)$, $\sigma_{\rm DVCS}(Q^2, W)$    & 6.5-80  &  30-140  & 2007 &   \cite{Aaron:2007ab}      \\  
H1 & $\sigma_{\rm DVCS}(Q^2), \sigma_{\rm DVCS}(W)$, $\sigma_{\rm DVCS}(Q^2, W)$   & 6.5-80  &  30-140  & 2009 &   \cite{Aaron:2009ac}      \\  
ZEUS & $\sigma_{\rm DVCS}(Q^2), \sigma_{\rm DVCS}(W)$, $\sigma_{\rm DVCS}(Q^2, W)$ & 5-100   &  40-140  & 2003 &  \cite{Chekanov:2003ya}    \\  
ZEUS & $\sigma_{\rm DVCS}(Q^2), \sigma_{\rm DVCS}(W)$, $\sigma_{\rm DVCS}(Q^2, W)$ & 1.5-100 &  40-170  & 2008 &  \cite{Chekanov:2008vy}    \\   \hline 
\hline
\end{tabular}
\caption{\small Overview of DVCS on proton experiments at HERA collider used in this study. 
The observable $\sigma_{\rm DVCS}$ is the cross section for the sub-process $\gamma^{\ast} p \rightarrow \gamma p$. } \label{table:DataSet} 
\end{table*}

\subsection{DVCS cross section in NLO pQCD vs. HERA data}

\begin{figure*}[htb]
	\begin{center}
		\resizebox{0.52\textwidth}{!}{\includegraphics{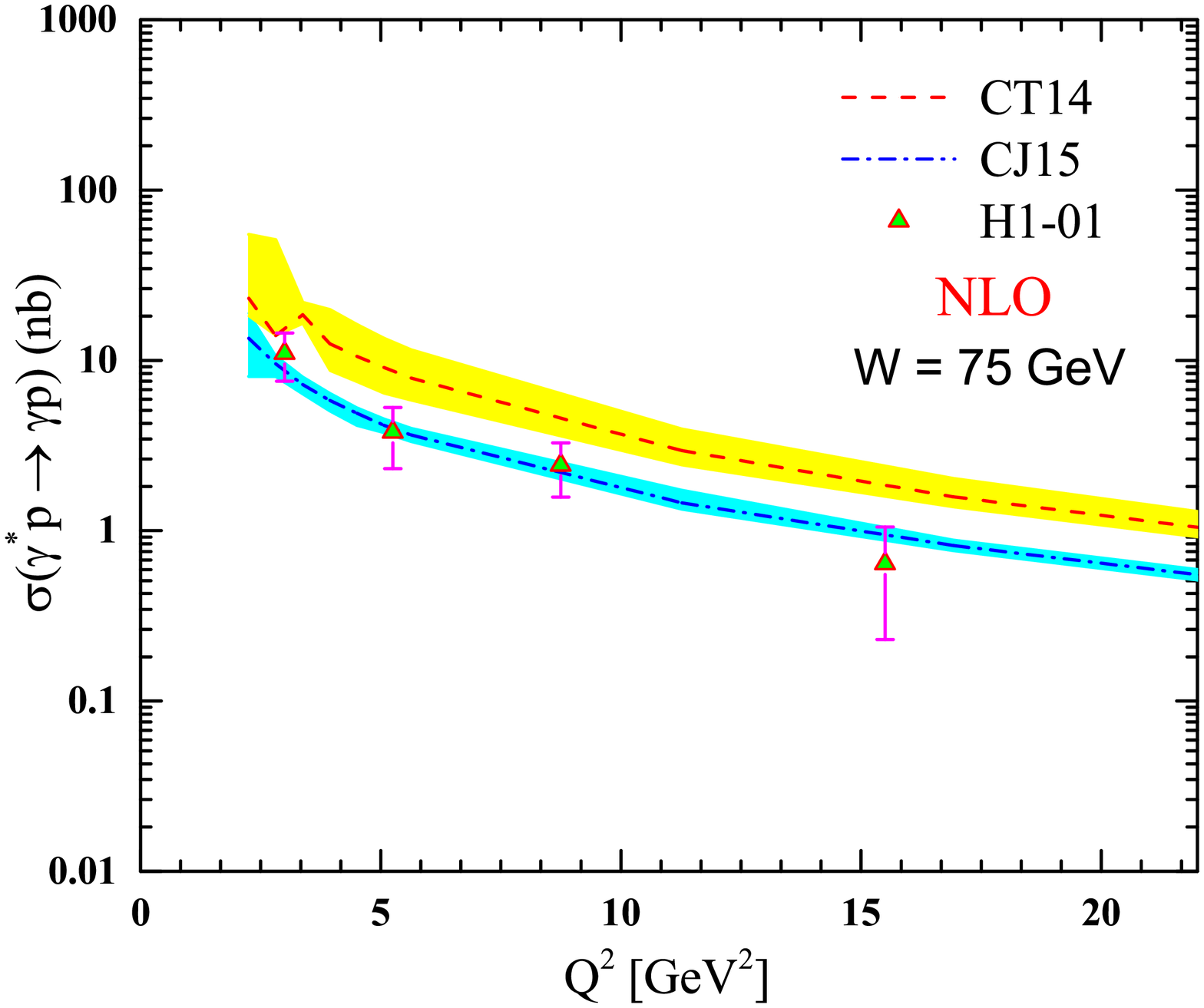}}   
		\hspace{-10mm}
		\resizebox{0.52\textwidth}{!}{\includegraphics{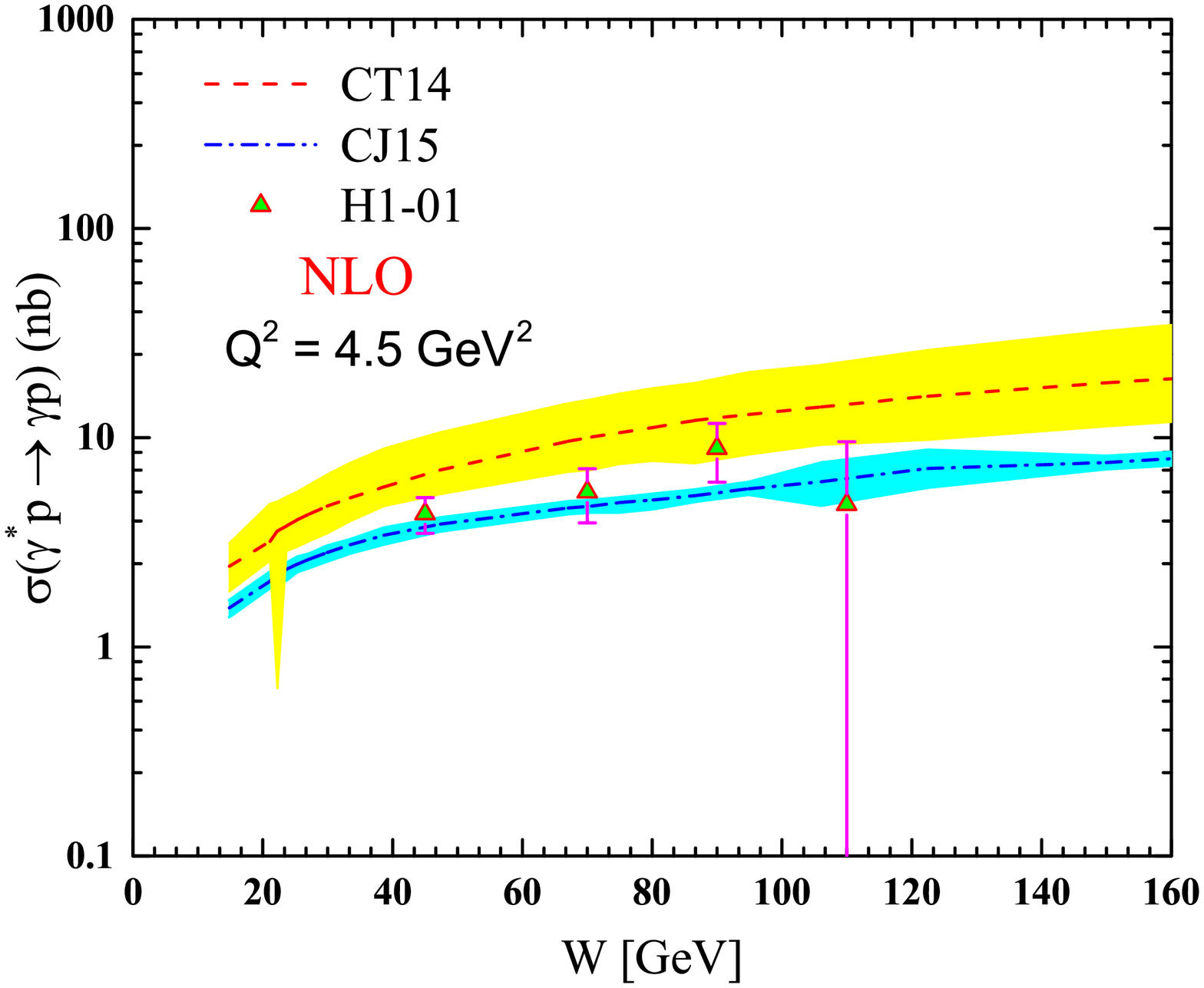}}   
\caption{(Color online) The DVCS $\gamma^{\ast} p \rightarrow \gamma p$ cross section as a function of $Q^2$ (left) and $W$ (right). The 2001 H1 data~\cite{Adloff:2001cn}, where the statistical and systematical errors are added in quadrature, is compared to
		our NLO pQCD results based on the input of Eq.~(\ref{eq:input}) and CT14~\cite{Dulat:2015mca} and 
		CJ15~\cite{Accardi:2016qay} PDFs. The shadowed bands represent the uncertainty of the corresponding PDFs.}
	 \label{fig:H1-2001}
	\end{center}
\end{figure*}

\begin{figure*}[htb]
	\begin{center}
		\resizebox{0.52\textwidth}{!}{\includegraphics{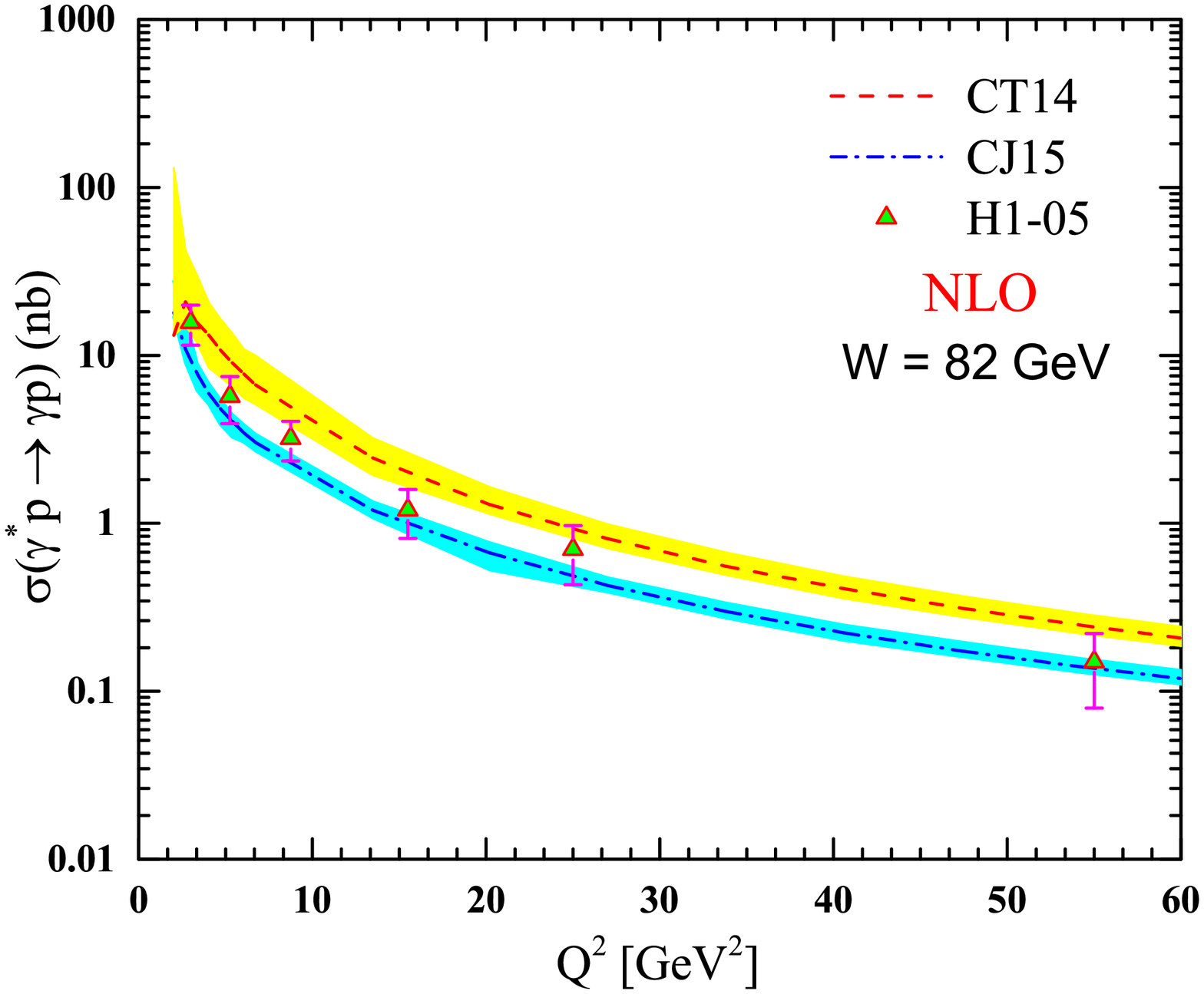}}   
		\hspace{-10mm}
		\resizebox{0.52\textwidth}{!}{\includegraphics{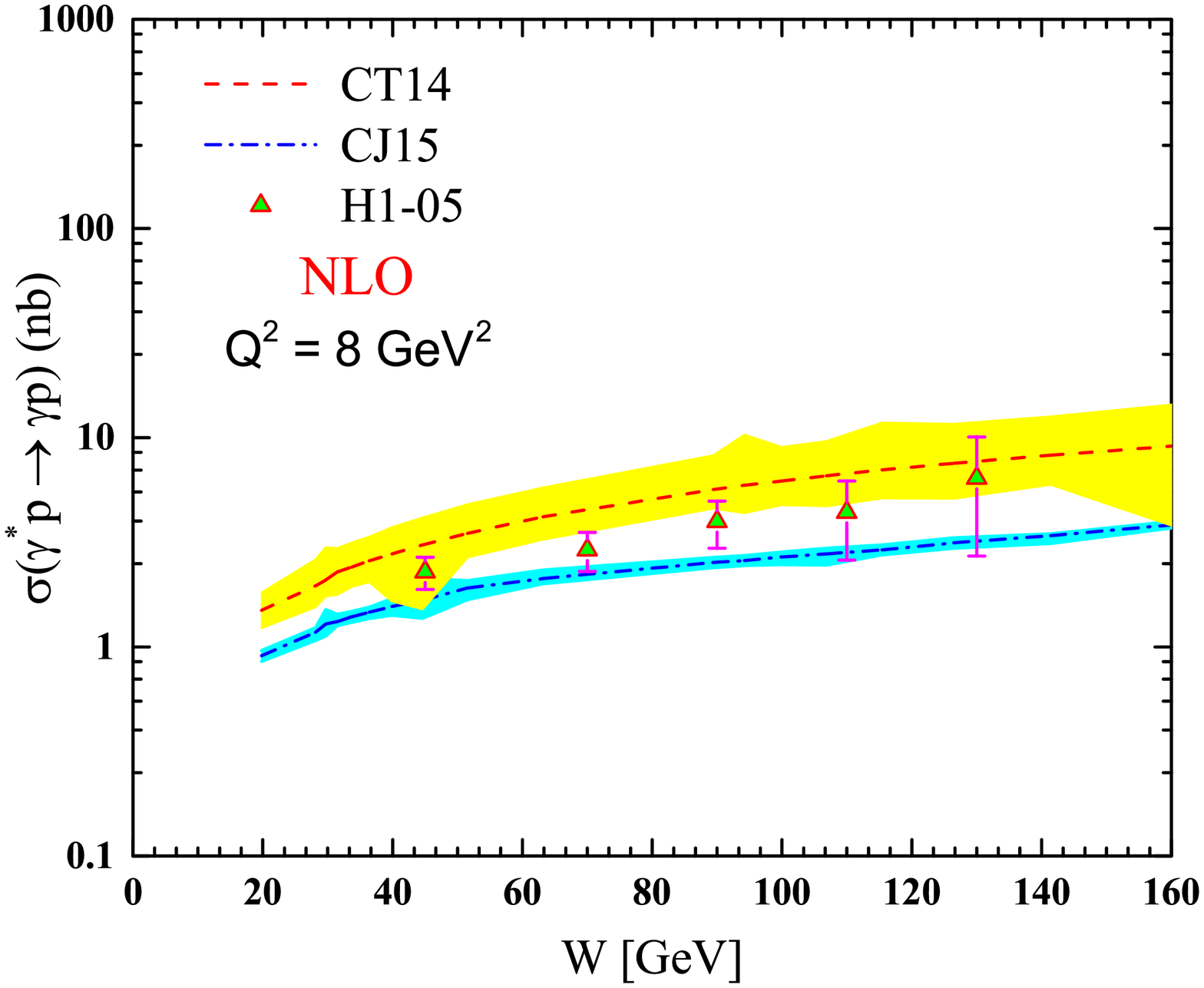}}   
		\caption{(Color online) 
		The DVCS $\gamma^{\ast} p \rightarrow \gamma p$ cross section as a function of $Q^2$ (left) and $W$ (right).
		Our NLO pQCD results are compared to the 2005 H1 data~\cite{Aktas:2005ty}, see details in Fig.~\ref{fig:H1-2001}.}
		\label{fig:H1-2005}
	\end{center}
\end{figure*}

\begin{figure*}[htb]
	\begin{center}
		\resizebox{0.52\textwidth}{!}{\includegraphics{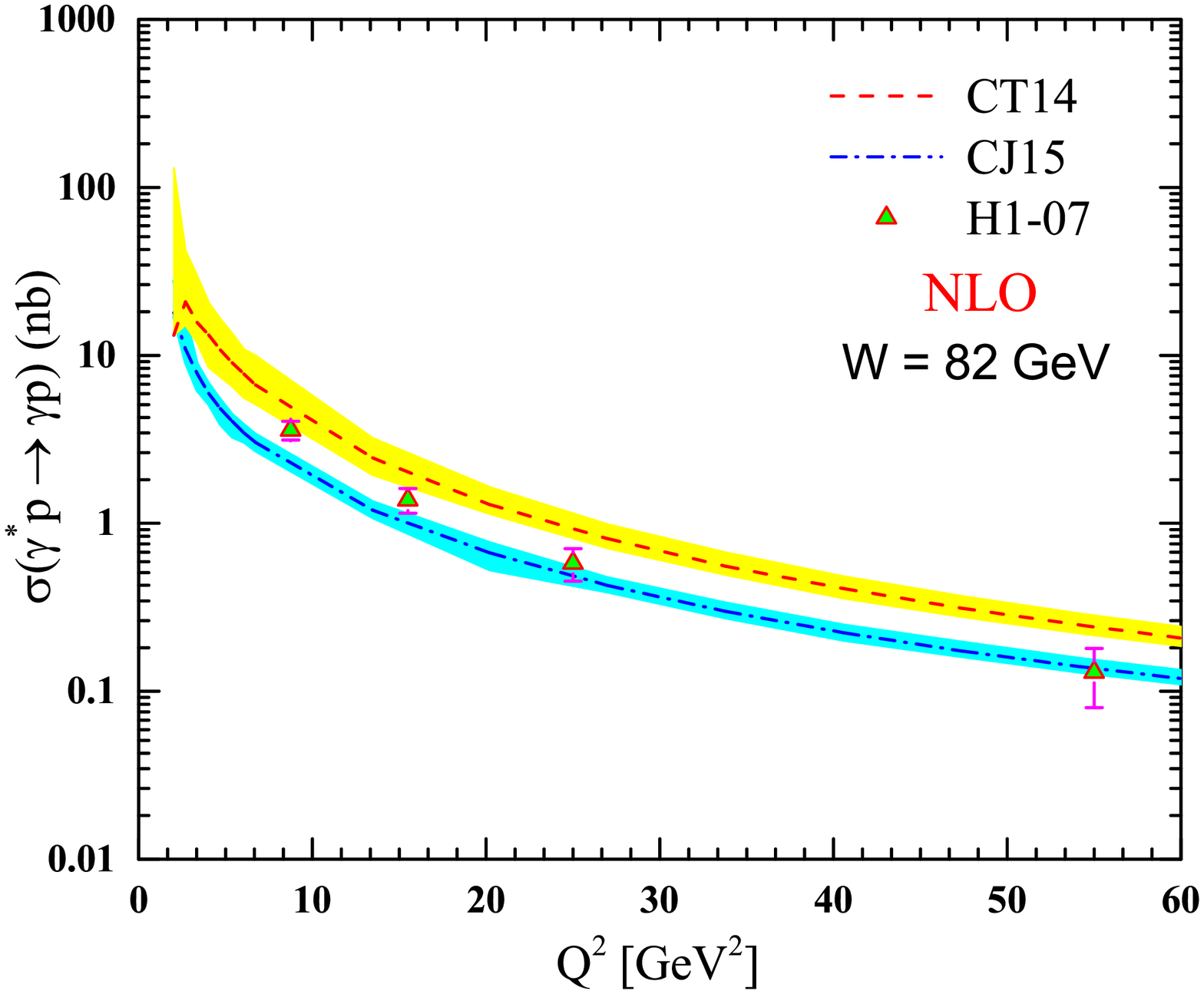}}   
		\hspace{-10mm}
		\resizebox{0.52\textwidth}{!}{\includegraphics{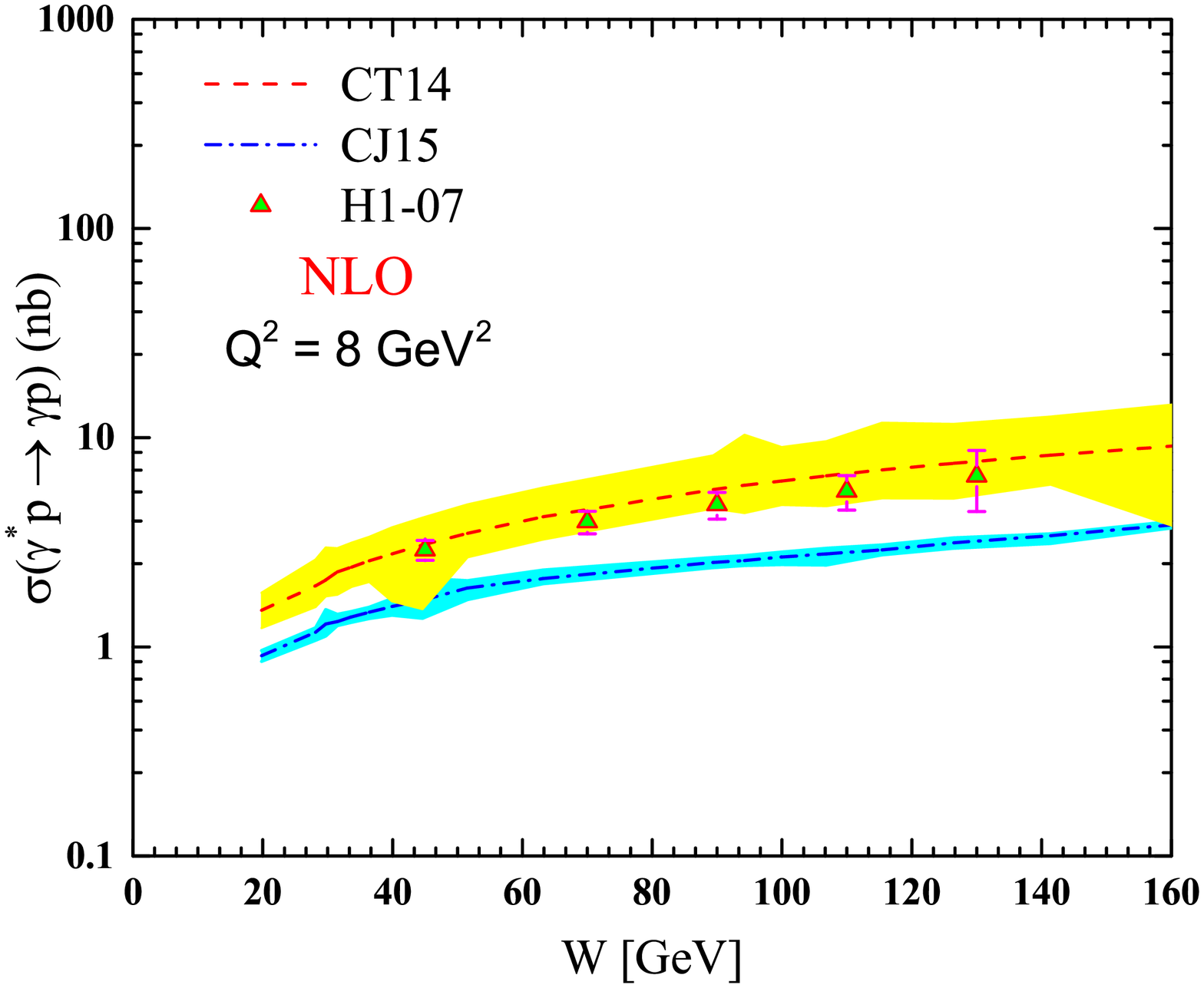}}   
		\caption{(Color online) 
			The DVCS $\gamma^{\ast} p \rightarrow \gamma p$ cross section as a function of $Q^2$ (left) and $W$ (right).
			Our NLO pQCD results are compared to the 2007 H1 data~\cite{Aaron:2007ab}, see details in Fig.~\ref{fig:H1-2001}.}
		\label{fig:H1-2007}
	\end{center}
\end{figure*}

\begin{figure*}[htb]
	\begin{center}
		\resizebox{0.52\textwidth}{!}{\includegraphics{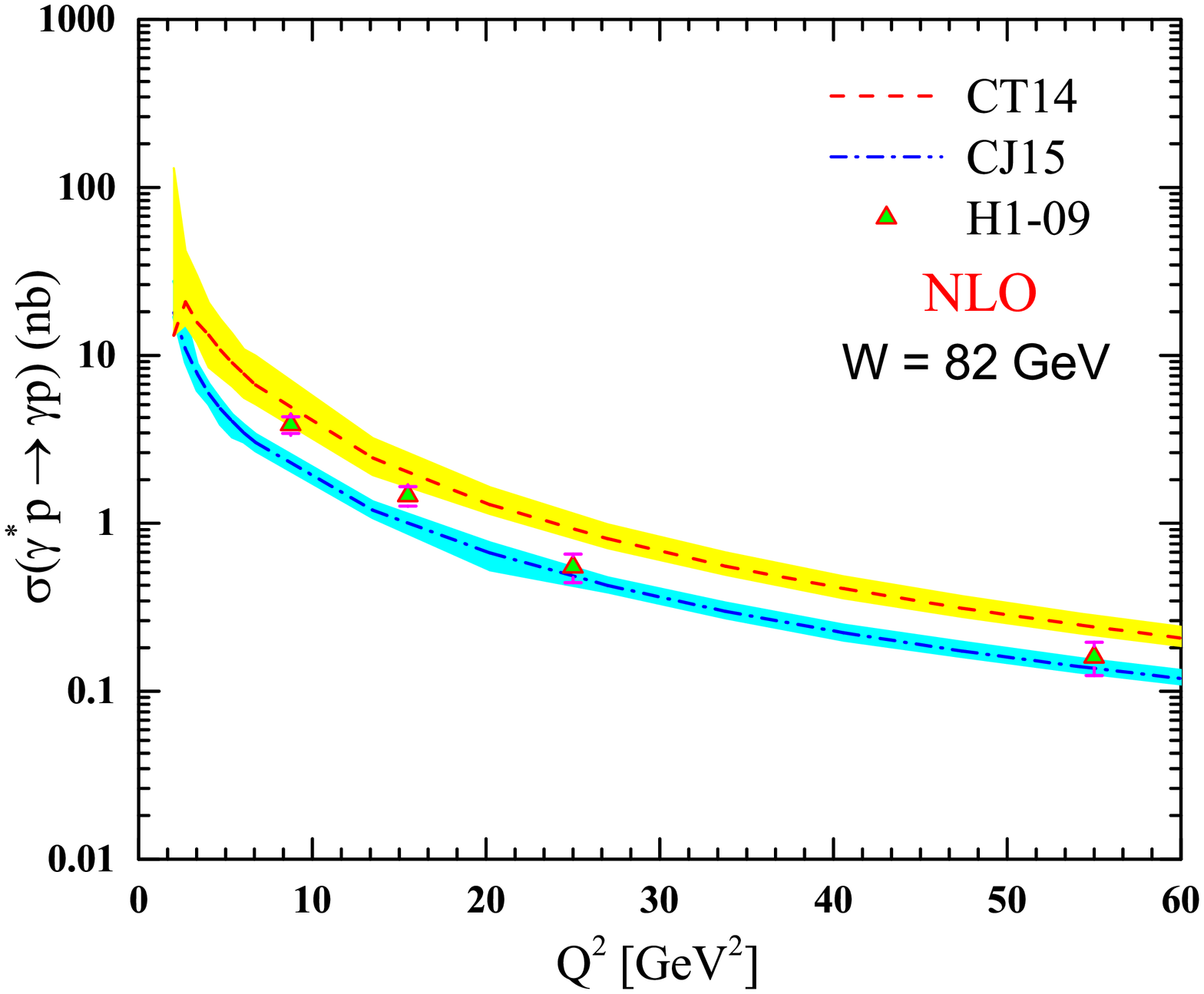}}   
		\hspace{-10mm}
		\resizebox{0.52\textwidth}{!}{\includegraphics{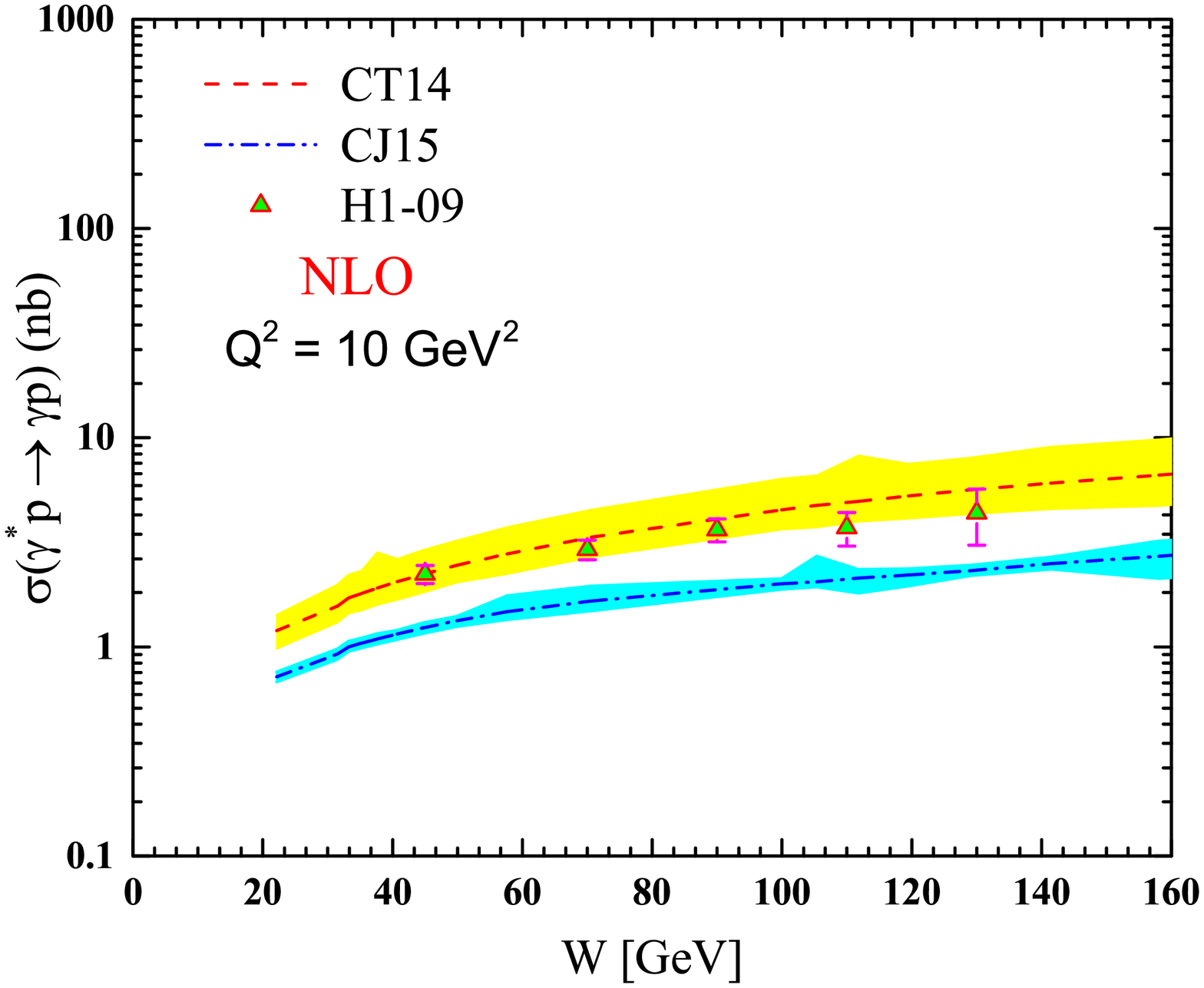}}   
		\caption{(Color online) 
		The DVCS $\gamma^{\ast} p \rightarrow \gamma p$ cross section as a function of $Q^2$ (left) and $W$ (right).
		Our NLO pQCD results are compared to the 2009 H1 data~\cite{Aaron:2009ac}, see details in Fig.~\ref{fig:H1-2001}.}		
	   \label{fig:H1-2009}
		\end{center}
\end{figure*}

\begin{figure*}[htb]
	\begin{center}
		\resizebox{0.52\textwidth}{!}{\includegraphics{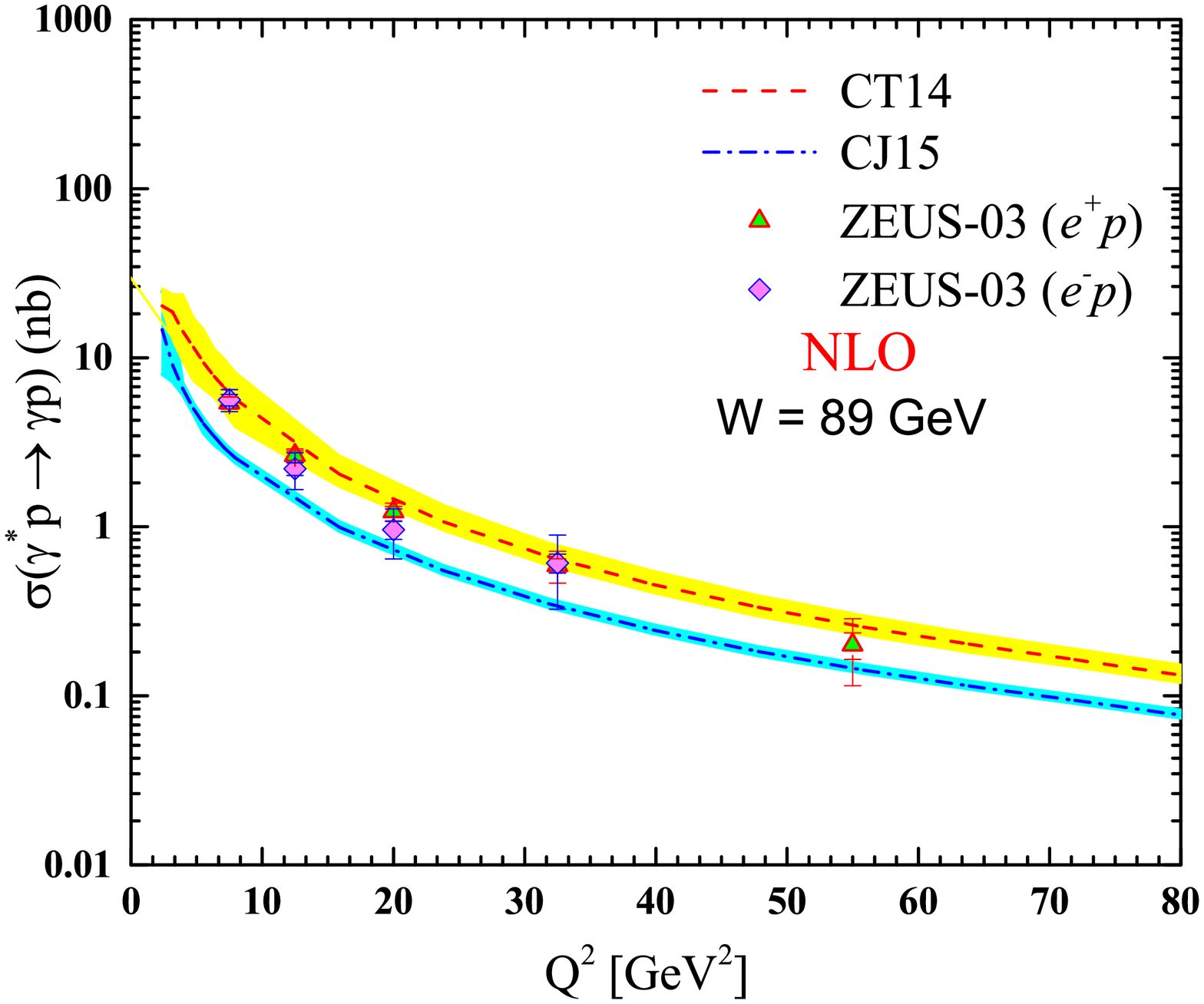}}   
		\hspace{-10mm}
		\resizebox{0.52\textwidth}{!}{\includegraphics{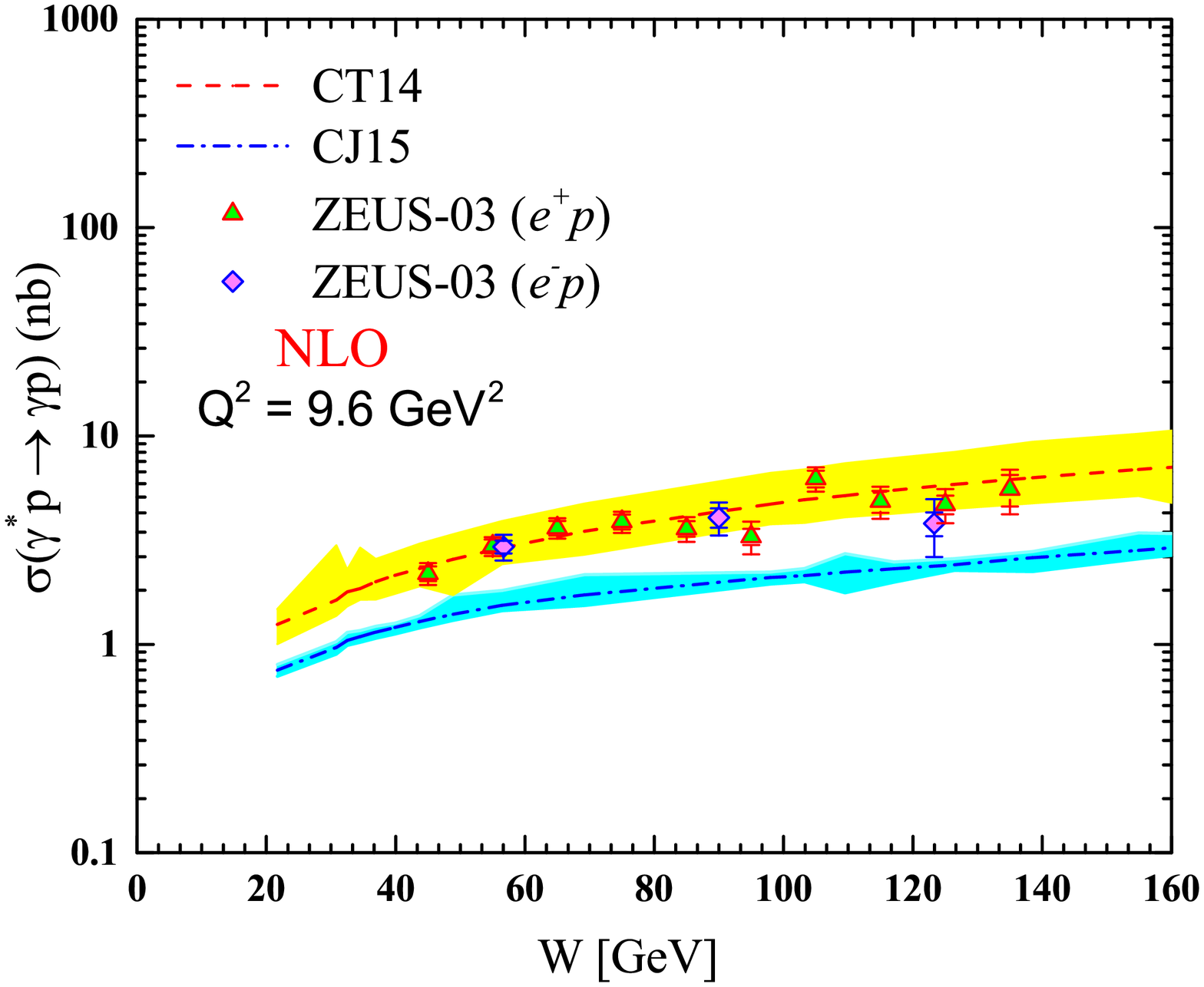}}   
		\caption{(Color online) 
			The DVCS $\gamma^{\ast} p \rightarrow \gamma p$ cross section as a function of $Q^2$ (left) and $W$ (right).
			Our NLO pQCD results are compared to the 2003 ZEUS data~\cite{Chekanov:2003ya}, see details in Fig.~\ref{fig:H1-2001}. The inner error bars represent the statistical errors, and the outer error bars the statistical and systematic errors added in quadrature.}
		\label{fig:ZEUS-2003}
	\end{center}
\end{figure*}

\begin{figure*}[htb]
	\begin{center}
		\resizebox{0.52\textwidth}{!}{\includegraphics{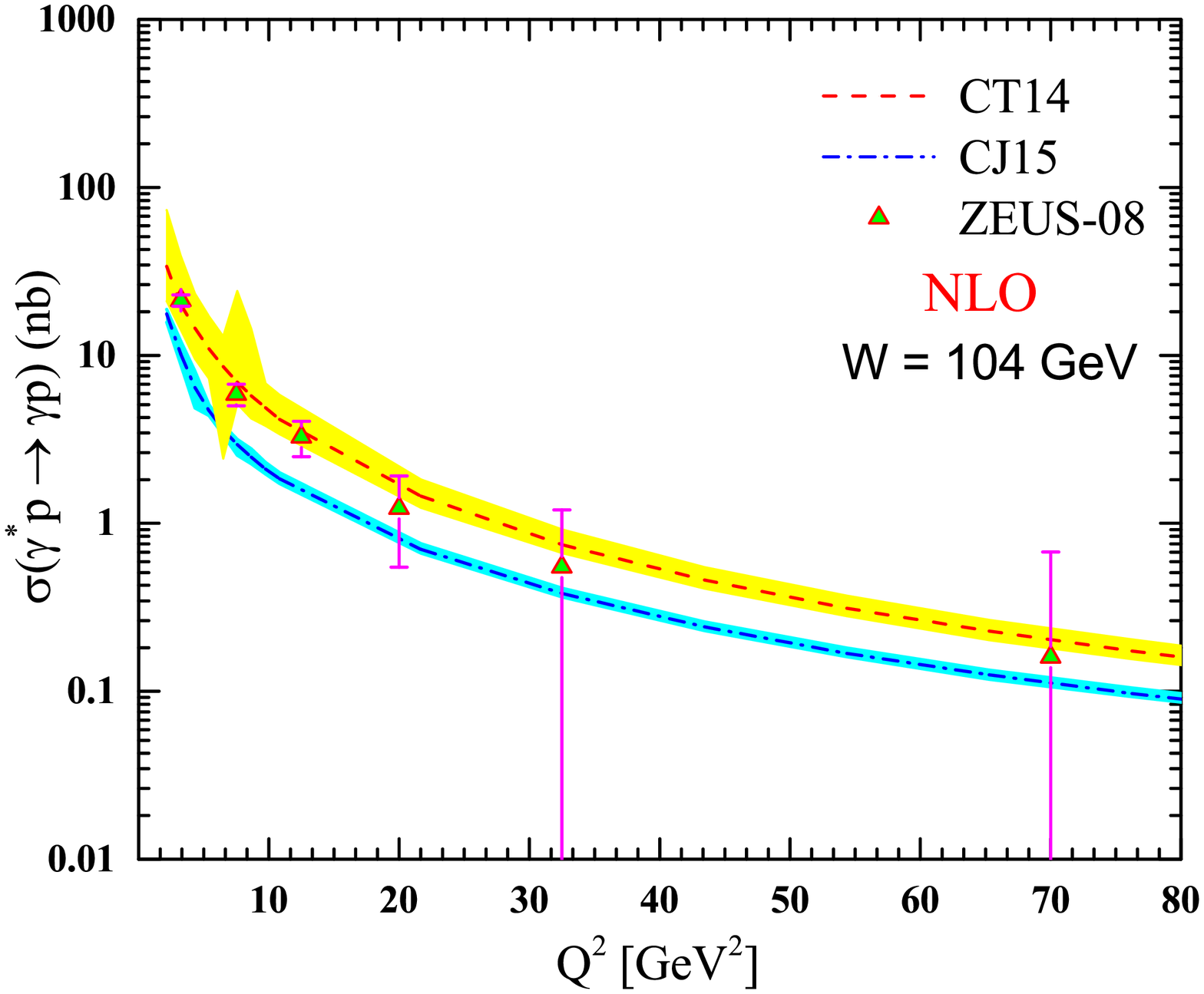}}   
		\hspace{-10mm}
		\resizebox{0.52\textwidth}{!}{\includegraphics{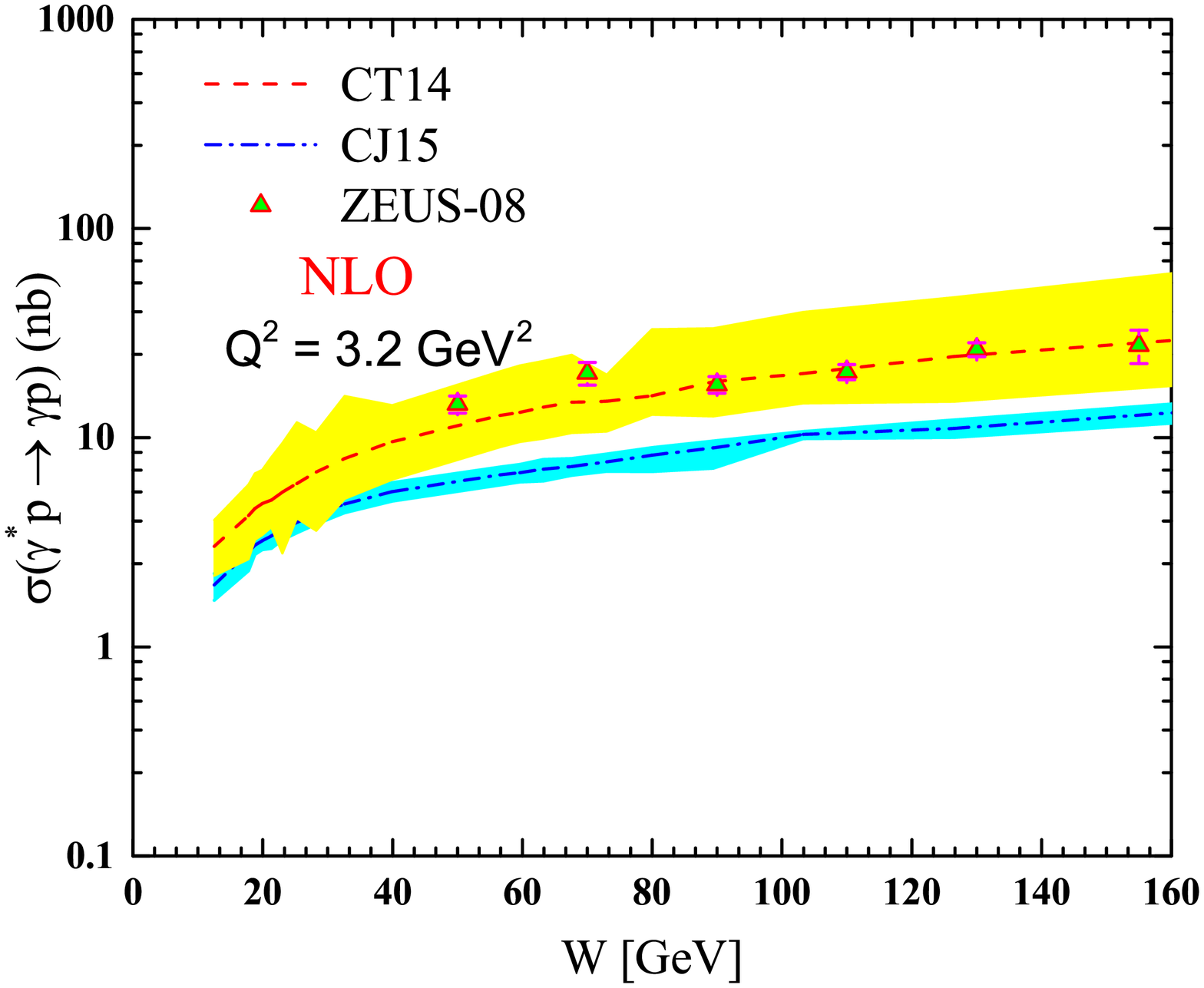}}   
		\caption{(Color online) 
		The DVCS $\gamma^{\ast} p \rightarrow \gamma p$ cross section as a function of $Q^2$ (left) and $W$ (right).
		Our NLO pQCD results are compared to the 2008 ZEUS data~\cite{Chekanov:2008vy}, see details in Fig.~\ref{fig:H1-2001}.}
	\label{fig:ZEUS-2008}
	\end{center}
\end{figure*}

Using our model for the singlet quark and gluon GPDs of the proton (see Sec.~\ref{sec:Modeling-of-GPDs}), we make predictions for the DVCS cross section in NLO
perturbative QCD. Our results are presented in Figs.~\ref{fig:H1-2001}, \ref{fig:H1-2005}, \ref{fig:H1-2007}, \ref{fig:H1-2009}, \ref{fig:ZEUS-2003} and \ref{fig:ZEUS-2008},
where they are compared to the available HERA data of the H1~\cite{Adloff:2001cn,Aktas:2005ty,Aaron:2007ab,Aaron:2009ac} 
and ZEUS~\cite{Chekanov:2003ya,Chekanov:2008vy} measurements (see Table~\ref{table:DataSet}).
The error bars the statistical and systematic uncertainties added in quadrature. The bands associated with CJ15 and CT14 prediction correspond to the uncertainty of the respective PDFs.  

One can see from these figures that within experimental and theoretical uncertainties, the input GPD model based on the 
CJ15 fit provides a good description of the H1-2001, H1-2005, H1-2007 and H1-2009 data  ($Q^2$ dependence only 
for the two latter data sets), while the model based on
the CT14 fit tends to  somewhat overestimate the cross section normalization
(it describes well the $W$ dependence of the H1-2005, H1-2007 and H1-2009 data).
At the same time, the CT14 parametrization leads to a very good description of the ZEUS data.
These results clearly show that for some selected PDF sets,  such as, e.g., the CJ15 and CT14 fits, the AJM GPD 
model of~\cite{Freund:2002qf} together with NLO pQCD calculations describes well the high-energy DVCS cross section.

\begin{figure}[htb]
\begin{center}
\resizebox{0.50\textwidth}{!}{\includegraphics{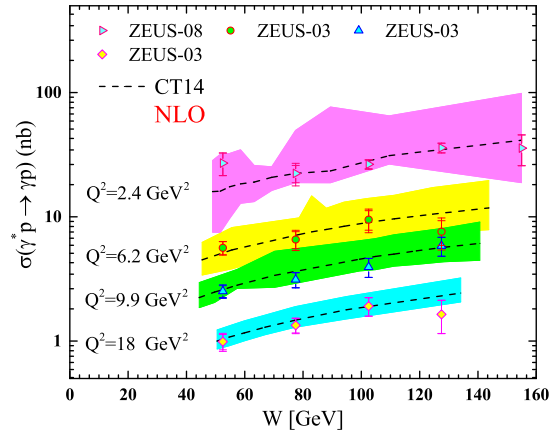}}
\caption{ (Color online) The DVCS $\gamma^{\ast} p \rightarrow \gamma p$ cross section as a function of $W$ for 
selected values of $Q^2 = 2.4$, $6.2$, $9.9$, and $18$ GeV$^2$. The NLO pQCD predictions based on the GPD model of Eq.~(\ref{eq:input}) along with the CT14~\cite{Dulat:2015mca} PDFs 
are compared to the 2003 and 2008 ZEUS data~\cite{Chekanov:2003ya,Chekanov:2008vy}. }
\label{fig:ZEUS-2003-2008-Q2-Evolution}
\end{center}
\end{figure}

In order to study effects of the NLO DGLAP evolution on GPDs, a detailed comparison of our obtained results with the 
DVCS $\gamma^{\ast} p \rightarrow \gamma p$ cross section is shown in Fig.~\ref{fig:ZEUS-2003-2008-Q2-Evolution}. 
In this figure, we show the DVCS cross section as a function of $W$ for some selected values of $Q^2 = 2.4, 6.2, 9.9$ and 18 GeV$^2$. 
Our NLO pQCD predictions are based on the CT14~\cite{Dulat:2015mca} PDFs; the experimental points are the 2003 and 2008 ZEUS data~\cite{Chekanov:2003ya,Chekanov:2008vy}. 
The inner error bars represent the statistical, and the full error bars the quadratic sum of the statistical and systematic uncertainties.
One can see that a very good agreement between our predictions and ZEUS data is achieved for a wide range of $Q^2$ and $W$.
It illustrates an important role of the $Q^2$-dependence of the quark and gluon GPD $H$ for the successful description
of the HERA data, which spans a wide range of $Q^2$ and $W$.

In summary, we observe very good overall agreement between our NLO pQCD predictions and most of the H1 and ZEUS data. 
It warrants the application of our framework to forthcoming and planned DVCS measurements at high energies, such as, e.g.
at COMPASS at CERN~\cite{Gautheron:2010wva}, an Electron-Ion Collider (EIC)~\cite{Aschenauer:2013hhw}, and the Large Hadron-Electron Collider (LHeC)~\cite{AbelleiraFernandez:2012cc} or Future Circular Collider
(FCC-he). 

Moreover, using our GPD model as a baseline, 
one can perform a global fit all available H1 and ZEUS DVCS data (cross section and its asymmetries) 
as well as the data from other DVCS experiments with fixed proton targets (HERMES, JLab) (the latter will require extension of our model to the remaining GPDs).
It is worth mentioning here that all known constraints on GPDs presented in Sec.~\ref{sec:Input-GPDs} 
cause the reduction of flexibility of choosing a proper GPDs functional from. 
The success of global fits existing in the literature (see their brief discussion in Introduction) as well as any future attempts to global fitting procedures strongly depend on the choice of data sets and the GPDs functional form. Therefore, any advances both in theory and experiments in these regards are most welcome.

\clearpage

\section{Conclusions}\label{sec:Conclusions}

The DVCS process is the golden channel to access GPDs and potentially extract them from the experimental observables.
Taking advantage of the high invariant energy available in lepton-proton collisions at HERA, 
the H1 and ZEUS measured the DVCS cross section in a wide kinematic range and studied precisely its dependence on
$Q^2$, $W$, and $t$. These measurements covered the Bjorken $x$ range of $10^{-4} < x_B <10^{-2}$, 
where sea quarks and gluons dominate. These data sets provide valuable information for GPDs phenomenology 
and several groups have attempted to extract CFFs and GPDs using them.

In this work, we studied the effects of NLO $Q^2$ evolution of GPDs using a model for the singlet quark and gluon
GPDs at an initial evolution scale motivated by the aligned-jet model of photon--hadron interactions at high energies.
Quantifying the skewness and evolution effects by the GPD-to-PDF
ratios $r^S$ and $r^g$, we found that $r^S$ increases logarithmically slowly from $r^S=1$ 
at the input scale of $Q^2=1.69$ GeV$^2$ to $r^S=1.5-2$ at $Q^2=100$ GeV$^2$; in the gluon channel, 
$r^g \approx 1$ for the studied range of $Q^2$. This observation agrees with the results of the more sophisticated model of GPDs
based on conformal expansion~\cite{Kumericki:2009uq}.

Using the resulting GPDs, we calculated the DVCS cross section on the proton in NLO pQCD and compared it to the
available HERA data. We found that our simple physical model of input GPDs used in conjunction with two modern
parameterizations of proton PDFs (CJ15 and CT14) provides good description of the H1 and ZEUS data.
It demonstrates that our GPDs model is reliable and flexible enough to be used in fitting procedures using variety of data sets. 

%
\section*{Acknowledgments}

VG would like to thank M.~Strikman and M.~Diehl for useful discussions of the model of input GPDs used in this analysis.
HK acknowledges the University of Science and Technology of Mazandaran and School of Particles and Accelerators, Institute for Research in Fundamental Sciences (IPM) for financial support provided for this research.



\begin{thebibliography}{99}

\bibitem{Mueller:1998fv} 
D.~M\"uller, D.~Robaschik, B.~Geyer, F.-M.~Dittes and J.~Ho\v{r}ej\v{s}i,
Fortsch.\ Phys.\  {\bf 42}, 101 (1994)
doi:10.1002/prop.2190420202
[hep-ph/9812448].


\bibitem{Radyushkin:1997ki} 
A.~V.~Radyushkin,
Phys.\ Rev.\ D {\bf 56}, 5524 (1997)
doi:10.1103/PhysRevD.56.5524
[hep-ph/9704207].


\bibitem{Ji:1996nm} 
X.~D.~Ji,
Phys.\ Rev.\ D {\bf 55}, 7114 (1997)
doi:10.1103/PhysRevD.55.7114
[hep-ph/9609381].


\bibitem{Ji:1998pc} 
X.~D.~Ji,
J.\ Phys.\ G {\bf 24}, 1181 (1998)
doi:10.1088/0954-3899/24/7/002
[hep-ph/9807358].


\bibitem{Goeke:2001tz} 
K.~Goeke, M.~V.~Polyakov and M.~Vanderhaeghen,
Prog.\ Part.\ Nucl.\ Phys.\  {\bf 47}, 401 (2001)
doi:10.1016/S0146-6410(01)00158-2
[hep-ph/0106012].


\bibitem{Belitsky:2001ns} 
A.~V.~Belitsky, D.~Mueller and A.~Kirchner,
Nucl.\ Phys.\ B {\bf 629}, 323 (2002)
doi:10.1016/S0550-3213(02)00144-X
[hep-ph/0112108].


\bibitem{Diehl:2003ny} 
M.~Diehl,
Phys.\ Rept.\  {\bf 388}, 41 (2003)
doi:10.1016/j.physrep.2003.08.002, 10.3204/DESY-THESIS-2003-018
[hep-ph/0307382].


\bibitem{Belitsky:2005qn} 
A.~V.~Belitsky and A.~V.~Radyushkin,
Phys.\ Rept.\  {\bf 418}, 1 (2005)
doi:10.1016/j.physrep.2005.06.002
[hep-ph/0504030].


\bibitem{Collins:1998be} 
J.~C.~Collins and A.~Freund,
Phys.\ Rev.\ D {\bf 59}, 074009 (1999)
doi:10.1103/PhysRevD.59.074009
[hep-ph/9801262].


\bibitem{Collins:1996fb} 
J.~C.~Collins, L.~Frankfurt and M.~Strikman,
Phys.\ Rev.\ D {\bf 56}, 2982 (1997)
doi:10.1103/PhysRevD.56.2982
[hep-ph/9611433].


\bibitem{Ivanov:2004vd} 
D.~Y.~Ivanov, A.~Schafer, L.~Szymanowski and G.~Krasnikov,
Eur.\ Phys.\ J.\ C {\bf 34}, no. 3, 297 (2004)
Erratum: [Eur.\ Phys.\ J.\ C {\bf 75}, no. 2, 75 (2015)]
doi:10.1140/epjc/s2004-01712-x, 10.1140/epjc/s10052-015-3298-8
[hep-ph/0401131].


\bibitem{Jones:2015nna} 
S.~P.~Jones, A.~D.~Martin, M.~G.~Ryskin and T.~Teubner,
J.\ Phys.\ G {\bf 43}, no. 3, 035002 (2016)
doi:10.1088/0954-3899/43/3/035002
[arXiv:1507.06942 [hep-ph]].


\bibitem{Polyakov:2002yz} 
M.~V.~Polyakov,
Phys.\ Lett.\ B {\bf 555}, 57 (2003)
doi:10.1016/S0370-2693(03)00036-4
[hep-ph/0210165].


\bibitem{Hagler:2007xi} 
P.~Hagler {\it et al.} [LHPC Collaboration],
Phys.\ Rev.\ D {\bf 77}, 094502 (2008)
doi:10.1103/PhysRevD.77.094502
[arXiv:0705.4295 [hep-lat]].


\bibitem{Alexandrou:2013joa} 
C.~Alexandrou, M.~Constantinou, S.~Dinter, V.~Drach, K.~Jansen, C.~Kallidonis and G.~Koutsou,
Phys.\ Rev.\ D {\bf 88}, no. 1, 014509 (2013)
doi:10.1103/PhysRevD.88.014509
[arXiv:1303.5979 [hep-lat]].


\bibitem{Freund:2001hm} 
A.~Freund and M.~F.~McDermott,
Phys.\ Rev.\ D {\bf 65}, 091901 (2002)
doi:10.1103/PhysRevD.65.091901
[hep-ph/0106124].


\bibitem{Freund:2001rk} 
A.~Freund and M.~F.~McDermott,
Phys.\ Rev.\ D {\bf 65}, 074008 (2002)
doi:10.1103/PhysRevD.65.074008
[hep-ph/0106319].


\bibitem{Freund:2001hd} 
A.~Freund and M.~McDermott,
Eur.\ Phys.\ J.\ C {\bf 23}, 651 (2002)
doi:10.1007/s100520200928
[hep-ph/0111472].


\bibitem{Ji:1997gm} 
X.~D.~Ji, W.~Melnitchouk and X.~Song,
Phys.\ Rev.\ D {\bf 56}, 5511 (1997)
doi:10.1103/PhysRevD.56.5511
[hep-ph/9702379].


\bibitem{Petrov:1998kf} 
V.~Y.~Petrov, P.~V.~Pobylitsa, M.~V.~Polyakov, I.~Bornig, K.~Goeke and C.~Weiss,
Phys.\ Rev.\ D {\bf 57}, 4325 (1998)
doi:10.1103/PhysRevD.57.4325
[hep-ph/9710270].


\bibitem{Tiburzi:2001ta} 
B.~C.~Tiburzi and G.~A.~Miller,
Phys.\ Rev.\ C {\bf 64}, 065204 (2001)
doi:10.1103/PhysRevC.64.065204
[hep-ph/0104198].
B.~C.~Tiburzi and G.~A.~Miller,
Phys.\ Rev.\ D {\bf 65}, 074009 (2002)
doi:10.1103/PhysRevD.65.074009
[hep-ph/0109174].



\bibitem{Scopetta:2003et} 
S.~Scopetta and V.~Vento,
Phys.\ Rev.\ D {\bf 69}, 094004 (2004)
doi:10.1103/PhysRevD.69.094004
[hep-ph/0307150].



\bibitem{Tiburzi:2004mh} 
B.~C.~Tiburzi, W.~Detmold and G.~A.~Miller,
Phys.\ Rev.\ D {\bf 70}, 093008 (2004)
doi:10.1103/PhysRevD.70.093008
[hep-ph/0408365].



\bibitem{Mineo:2005qr} 
H.~Mineo, S.~N.~Yang, C.~Y.~Cheung and W.~Bentz,
Nucl.\ Phys.\ Proc.\ Suppl.\  {\bf 141}, 281 (2005)
doi:10.1016/j.nuclphysbps.2004.12.042
[hep-ph/0502017].



\bibitem{Pobylitsa:2002vw} 
P.~V.~Pobylitsa,
Phys.\ Rev.\ D {\bf 67}, 094012 (2003)
doi:10.1103/PhysRevD.67.094012
[hep-ph/0210238].


\bibitem{Freund:2002qf} 
A.~Freund, M.~McDermott and M.~Strikman,
Phys.\ Rev.\ D {\bf 67}, 036001 (2003)
doi:10.1103/PhysRevD.67.036001
[hep-ph/0208160].

  

\bibitem{Kumericki:2007sa} 
K.~Kumericki, D.~Mueller and K.~Passek-Kumericki,
Nucl.\ Phys.\ B {\bf 794}, 244 (2008)
doi:10.1016/j.nuclphysb.2007.10.029
[hep-ph/0703179 [HEP-PH]].

  

\bibitem{Kumericki:2009uq} 
K.~Kumeri?ki and D.~Mueller,
Nucl.\ Phys.\ B {\bf 841}, 1 (2010)
doi:10.1016/j.nuclphysb.2010.07.015
[arXiv:0904.0458 [hep-ph]].
  


\bibitem{Lautenschlager:2013uya} 
T.~Lautenschlager, D.~Muller and A.~Schaefer,
arXiv:1312.5493 [hep-ph].


\bibitem{Kumericki:2016ehc} 
K.~Kumericki, S.~Liuti and H.~Moutarde,
Eur.\ Phys.\ J.\ A {\bf 52}, no. 6, 157 (2016)
doi:10.1140/epja/i2016-16157-3
[arXiv:1602.02763 [hep-ph]].


\bibitem{Guidal:2008ie} 
M.~Guidal,
Eur.\ Phys.\ J.\ A {\bf 37}, 319 (2008)
Erratum: [Eur.\ Phys.\ J.\ A {\bf 40}, 119 (2009)]
doi:10.1140/epja/i2008-10630-6, 10.1140/epja/i2009-10748-y
[arXiv:0807.2355 [hep-ph]].


\bibitem{Guidal:2010ig} 
M.~Guidal,
Phys.\ Lett.\ B {\bf 689}, 156 (2010)
doi:10.1016/j.physletb.2010.04.053
[arXiv:1003.0307 [hep-ph]].
  

\bibitem{Guidal:2013rya} 
M.~Guidal, H.~Moutarde and M.~Vanderhaeghen,
Rept.\ Prog.\ Phys.\  {\bf 76}, 066202 (2013)
doi:10.1088/0034-4885/76/6/066202
[arXiv:1303.6600 [hep-ph]].
  

\bibitem{Berthou:2015oaw} 
B.~Berthou {\it et al.},
arXiv:1512.06174 [hep-ph].
  
\bibitem{Radyushkin:1998es} 
A.~V.~Radyushkin,
Phys.\ Rev.\ D {\bf 59}, 014030 (1999)
doi:10.1103/PhysRevD.59.014030
[hep-ph/9805342].

\bibitem{Radyushkin:1998bz} 
A.~V.~Radyushkin,
Phys.\ Lett.\ B {\bf 449}, 81 (1999)
doi:10.1016/S0370-2693(98)01584-6
[hep-ph/9810466].

\bibitem{Musatov:1999xp} 
I.~V.~Musatov and A.~V.~Radyushkin,
Phys.\ Rev.\ D {\bf 61}, 074027 (2000)
doi:10.1103/PhysRevD.61.074027
[hep-ph/9905376].


\bibitem{Guidal:2004nd} 
M.~Guidal, M.~V.~Polyakov, A.~V.~Radyushkin and M.~Vanderhaeghen,
Phys.\ Rev.\ D {\bf 72}, 054013 (2005)
doi:10.1103/PhysRevD.72.054013
[hep-ph/0410251].

\bibitem{Radyushkin:2013hca} 
A.~V.~Radyushkin,
Phys.\ Rev.\ D {\bf 87}, no. 9, 096017 (2013)
doi:10.1103/PhysRevD.87.096017
[arXiv:1304.2682 [hep-ph]].


\bibitem{Kumericki:2011rz} 
K.~Kumericki, D.~Mueller and A.~Schafer,
JHEP {\bf 1107}, 073 (2011)
doi:10.1007/JHEP07(2011)073
[arXiv:1106.2808 [hep-ph]].

\bibitem{Aschenauer:2013hhw} 
E.~C.~Aschenauer, S.~Fazio, K.~Kumericki and D.~Mueller,
JHEP {\bf 1309}, 093 (2013)
doi:10.1007/JHEP09(2013)093
[arXiv:1304.0077 [hep-ph]].
  
\bibitem{Bjorken:1973gc} 
J.~D.~Bjorken and J.~B.~Kogut,
Phys.\ Rev.\ D {\bf 8}, 1341 (1973).
doi:10.1103/PhysRevD.8.1341

\bibitem{Frankfurt:1988nt} 
L.~L.~Frankfurt and M.~I.~Strikman,
Phys.\ Rept.\  {\bf 160}, 235 (1988).
doi:10.1016/0370-1573(88)90179-2

\bibitem{Frankfurt:2013ria} 
L.~Frankfurt and M.~Strikman,
doi:10.1142/9789814425810\_0014
arXiv:1304.4308 [hep-ph].
  
\bibitem{Frankfurt:1997at} 
L.~L.~Frankfurt, A.~Freund and M.~Strikman,
Phys.\ Rev.\ D {\bf 58}, 114001 (1998)
Erratum: [Phys.\ Rev.\ D {\bf 59}, 119901 (1999)]
doi:10.1103/PhysRevD.58.114001, 10.1103/PhysRevD.59.119901
[hep-ph/9710356].


\bibitem{Schoeffel:2007dt} 
L.~Schoeffel,
Phys.\ Lett.\ B {\bf 658}, 33 (2007)
doi:10.1016/j.physletb.2007.10.036
[arXiv:0706.3488 [hep-ph]].

\bibitem{Guichon:1998xv} 
P.~A.~M.~Guichon and M.~Vanderhaeghen,
Prog.\ Part.\ Nucl.\ Phys.\  {\bf 41}, 125 (1998)
doi:10.1016/S0146-6410(98)00056-8
[hep-ph/9806305].

\bibitem{Vanderhaeghen:1999xj} 
M.~Vanderhaeghen, P.~A.~M.~Guichon and M.~Guidal,
Phys.\ Rev.\ D {\bf 60}, 094017 (1999)
doi:10.1103/PhysRevD.60.094017
[hep-ph/9905372].

\bibitem{Diehl:2007jb} 
M.~Diehl and D.~Y.~Ivanov,
Eur.\ Phys.\ J.\ C {\bf 52}, 919 (2007)
doi:10.1140/epjc/s10052-007-0401-9
[arXiv:0707.0351 [hep-ph]].

\bibitem{GolecBiernat:1999ib} 
K.~J.~Golec-Biernat, A.~D.~Martin and M.~G.~Ryskin,
Phys.\ Lett.\ B {\bf 456}, 232 (1999)
doi:10.1016/S0370-2693(99)00504-3
[hep-ph/9903327].

\bibitem{Guzey:2005ec} 
V.~Guzey and M.~V.~Polyakov,
Eur.\ Phys.\ J.\ C {\bf 46}, 151 (2006)
doi:10.1140/epjc/s2006-02491-0
[hep-ph/0507183].

\bibitem{Guzey:2006xi} 
V.~Guzey and T.~Teckentrup,
Phys.\ Rev.\ D {\bf 74}, 054027 (2006)
doi:10.1103/PhysRevD.74.054027
[hep-ph/0607099].
V.~Guzey and T.~Teckentrup,
Phys.\ Rev.\ D {\bf 79}, 017501 (2009)
doi:10.1103/PhysRevD.79.017501
[arXiv:0810.3899 [hep-ph]].

\bibitem{Polyakov:1999gs} 
M.~V.~Polyakov and C.~Weiss,
Phys.\ Rev.\ D {\bf 60}, 114017 (1999)
doi:10.1103/PhysRevD.60.114017
[hep-ph/9902451].
 

\bibitem{Kivel:2000fg} 
N.~Kivel, M.~V.~Polyakov and M.~Vanderhaeghen,
Phys.\ Rev.\ D {\bf 63}, 114014 (2001)
doi:10.1103/PhysRevD.63.114014
[hep-ph/0012136].

\bibitem{Ball:2017nwa} 
R.~D.~Ball {\it et al.} [NNPDF Collaboration],
arXiv:1706.00428 [hep-ph].

\bibitem{Hou:2016nqm} 
T.~J.~Hou {\it et al.},
Phys.\ Rev.\ D {\bf 95}, no. 3, 034003 (2017)
doi:10.1103/PhysRevD.95.034003
[arXiv:1609.07968 [hep-ph]].

\bibitem{Harland-Lang:2014zoa} 
L.~A.~Harland-Lang, A.~D.~Martin, P.~Motylinski and R.~S.~Thorne,
Eur.\ Phys.\ J.\ C {\bf 75}, no. 5, 204 (2015)
doi:10.1140/epjc/s10052-015-3397-6
[arXiv:1412.3989 [hep-ph]].

\bibitem{Harland-Lang:2016yfn} 
L.~A.~Harland-Lang, A.~D.~Martin, P.~Motylinski and R.~S.~Thorne,
Eur.\ Phys.\ J.\ C {\bf 76}, no. 4, 186 (2016)
doi:10.1140/epjc/s10052-016-4020-1
[arXiv:1601.03413 [hep-ph]].


\bibitem{Frankfurt:1997ha} 
L.~Frankfurt, A.~Freund, V.~Guzey and M.~Strikman,
Phys.\ Lett.\ B {\bf 418}, 345 (1998)
Erratum: [Phys.\ Lett.\ B {\bf 429}, 414 (1998)]
doi:10.1016/S0370-2693(97)01152-0
[hep-ph/9703449].

\bibitem{Belitsky:1999hf} 
A.~V.~Belitsky, A.~Freund and D.~Mueller,
Nucl.\ Phys.\ B {\bf 574}, 347 (2000)
doi:10.1016/S0550-3213(00)00012-2
[hep-ph/9912379].

\bibitem{Belitsky:1999fu} 
A.~V.~Belitsky and D.~Mueller,
Phys.\ Lett.\ B {\bf 464}, 249 (1999)
doi:10.1016/S0370-2693(99)01003-5
[hep-ph/9906409].

\bibitem{Belitsky:1998vj} 
A.~V.~Belitsky and D.~Mueller,
Nucl.\ Phys.\ B {\bf 527}, 207 (1998)
doi:10.1016/S0550-3213(98)00310-1
[hep-ph/9802411].

\bibitem{Belitsky:1998gc} 
A.~V.~Belitsky and D.~Mueller,
Nucl.\ Phys.\ B {\bf 537}, 397 (1999)
doi:10.1016/S0550-3213(98)00677-4
[hep-ph/9804379].

\bibitem{Belitsky:1999gu} 
A.~V.~Belitsky, D.~Mueller and A.~Freund,
Phys.\ Lett.\ B {\bf 461}, 270 (1999)
doi:10.1016/S0370-2693(99)00837-0
[hep-ph/9904477].

\bibitem{Dulat:2015mca} 
S.~Dulat {\it et al.},
Phys.\ Rev.\ D {\bf 93}, no. 3, 033006 (2016)
doi:10.1103/PhysRevD.93.033006
[arXiv:1506.07443 [hep-ph]].

\bibitem{Accardi:2016qay} 
A.~Accardi, L.~T.~Brady, W.~Melnitchouk, J.~F.~Owens and N.~Sato,
Phys.\ Rev.\ D {\bf 93}, no. 11, 114017 (2016)
doi:10.1103/PhysRevD.93.114017
[arXiv:1602.03154 [hep-ph]].


\bibitem{Pumplin:2001ct} 
  J.~Pumplin, D.~Stump, R.~Brock, D.~Casey, J.~Huston, J.~Kalk, H.~L.~Lai and W.~K.~Tung,
  Phys.\ Rev.\ D {\bf 65}, 014013 (2001)
  doi:10.1103/PhysRevD.65.014013
  [hep-ph/0101032].
  
\bibitem{Nadolsky:2008zw} 
  P.~M.~Nadolsky, H.~L.~Lai, Q.~H.~Cao, J.~Huston, J.~Pumplin, D.~Stump, W.~K.~Tung and C.-P.~Yuan,
  Phys.\ Rev.\ D {\bf 78}, 013004 (2008)
  doi:10.1103/PhysRevD.78.013004
  [arXiv:0802.0007 [hep-ph]].
  
  
\bibitem{Accardi:2016ndt} 
  A.~Accardi {\it et al.},
  Eur.\ Phys.\ J.\ C {\bf 76}, no. 8, 471 (2016)
  doi:10.1140/epjc/s10052-016-4285-4
  [arXiv:1603.08906 [hep-ph]].
  
\bibitem{Mueller:2005nz} 
D.~Mueller,
Phys.\ Lett.\ B {\bf 634}, 227 (2006)
doi:10.1016/j.physletb.2006.01.036
[hep-ph/0510109].

\bibitem{Kumericki:2006xx} 
K.~Kumericki, D.~Mueller, K.~Passek-Kumericki and A.~Schafer,
Phys.\ Lett.\ B {\bf 648}, 186 (2007)
doi:10.1016/j.physletb.2007.02.071
[hep-ph/0605237].

\bibitem{Kroll:2012sm} 
P.~Kroll, H.~Moutarde and F.~Sabatie,
Eur.\ Phys.\ J.\ C {\bf 73}, no. 1, 2278 (2013)
doi:10.1140/epjc/s10052-013-2278-0
[arXiv:1210.6975 [hep-ph]].

\bibitem{Freund:2003qs} 
A.~Freund,
Phys.\ Rev.\ D {\bf 68}, 096006 (2003)
doi:10.1103/PhysRevD.68.096006
[hep-ph/0306012].



\bibitem{Adloff:2001cn} 
C.~Adloff {\it et al.} [H1 Collaboration],
Phys.\ Lett.\ B {\bf 517}, 47 (2001)
doi:10.1016/S0370-2693(01)00939-X
[hep-ex/0107005].

\bibitem{Aktas:2005ty} 
A.~Aktas {\it et al.} [H1 Collaboration],
Eur.\ Phys.\ J.\ C {\bf 44}, 1 (2005)
doi:10.1140/epjc/s2005-02345-3
[hep-ex/0505061].

\bibitem{Aaron:2007ab} 
F.~D.~Aaron {\it et al.} [H1 Collaboration],
Phys.\ Lett.\ B {\bf 659}, 796 (2008)
doi:10.1016/j.physletb.2007.11.093
[arXiv:0709.4114 [hep-ex]].

\bibitem{Aaron:2009ac} 
F.~D.~Aaron {\it et al.} [H1 Collaboration],
Phys.\ Lett.\ B {\bf 681}, 391 (2009)
doi:10.1016/j.physletb.2009.10.035
[arXiv:0907.5289 [hep-ex]].

\bibitem{Chekanov:2003ya} 
S.~Chekanov {\it et al.} [ZEUS Collaboration],
Phys.\ Lett.\ B {\bf 573}, 46 (2003)
doi:10.1016/j.physletb.2003.08.048
[hep-ex/0305028].

\bibitem{Chekanov:2008vy} 
S.~Chekanov {\it et al.} [ZEUS Collaboration],
JHEP {\bf 0905}, 108 (2009)
doi:10.1088/1126-6708/2009/05/108
[arXiv:0812.2517 [hep-ex]].

\bibitem{Gautheron:2010wva} 
F.~Gautheron {\it et al.} [COMPASS Collaboration],
SPSC-P-340, CERN-SPSC-2010-014.

\bibitem{AbelleiraFernandez:2012cc}
  J.~L.~Abelleira Fernandez {\it et al.} [LHeC Study Group],
  J.\ Phys.\ G {\bf 39} (2012) 075001
  doi:10.1088/0954-3899/39/7/075001
  [arXiv:1206.2913 [physics.acc-ph]].

\end{thebibliography}
\end{document}